\documentclass[structabstract]{aa} 

\usepackage{graphics}
\usepackage{graphicx}
\usepackage{amsmath}
\usepackage{amssymb}

\usepackage{listings}
\usepackage{color}
\usepackage{subfigure}
\usepackage{natbib}
\usepackage{txfonts}
\usepackage[T1]{fontenc}
\usepackage[latin1]{inputenc}

\bibpunct{(}{)}{;}{a}{}{,} 

\author{Daniel Johansson\inst{1}
\and Cathy Horellou\inst{1}
\and Martin~W.~Sommer\inst{2}
\and Kaustuv Basu\inst{3,2}
\and Frank Bertoldi\inst{2}
\and Mark Birkinshaw\inst{4}
\and Katy Lancaster\inst{4}
\and Omar Lopez-Cruz\inst{5}
\and Hernan Quintana\inst{6}
}

\institute{Onsala Space Observatory, Chalmers University of
  Technology, SE-439 92 Onsala, Sweden \and Argelander-Institut f\"ur
  Astronomie, Auf dem H\"ugel 71, D-53121 Bonn, Germany \and
  Max-Planck Institut f\"ur Radioastronomie, Auf dem H\"ugel 69,
  D-53121 Bonn, Germany \and Department of Physics, University of
  Bristol, Tyndall Avenue, Bristol BS8 1TL, United Kingdom \and
  Instituto Nacional de Astrof\'isica, Optica y Electr\'onica (INAOE),
  Tonantzintla, Puebla 72840, Mexico \and Departamento de Astronom\'ia
  y Astrof\'isica, Pontificia Universidad Cat\'olica de Chile, Casilla
  306, Santiago 22, Chile }

\title{Submillimeter galaxies behind the Bullet Cluster (1E~0657-56)}

\abstract {Clusters of galaxies are effective gravitational lenses
  able to magnify background galaxies and making it possible to probe
  the fainter part of the galaxy population.  Submillimeter galaxies,
  which are believed to be star-forming galaxies at typical redshifts
  of 2 to 3, are a major contaminant to the extended Sunyaev-Zeldovich
  (SZ) signal of galaxy clusters. For a proper quantification of the
  SZ signal the contribution of submillimeter galaxies needs to be
  quantified.} {The aims of this study are to identify submillimeter
  sources in the field of the Bullet Cluster (1E~0657-56), a massive cluster
  of galaxies at $z\simeq 0.3$, measure their flux densities at
  870~$\mu \mathrm{m}$, and search for counterparts at other wavelengths to
  constrain their properties.} {We carried out deep observations of
  the submillimeter continuum emission at $870~\mu$m using the Large
  APEX BOlometer CAmera (LABOCA) on the Atacama Pathfinder EXperiment
  (APEX) telescope. Several numerical techniques were used to quantify
  the noise properties of the data and extract sources.} {In total,
  seventeen sources were found. Thirteen of them lie in the central 10
  arcminutes of the map, which has a pixel sensitivity of 1.2~mJy per
  $22''$ beam. After correction for flux boosting and gravitational
  lensing, the number counts are consistent with published submm
  measurements. Nine of the sources have infrared
  counterparts in Spitzer maps. The strongest submm detection
  coincides with a source previously reported at other wavelengths, at
  an estimated redshift $z\simeq 2.7$. If the submm flux arises from
  two images of a galaxy magnified by a total factor of 75, as models
  have suggested, its intrinsic flux would be around 0.6~mJy,
  consistent with an intrinsic luminosity below $10^{12} L_\odot$.} {}

\date{Received / Accepted 19/02/2010 }


\titlerunning{Submillimeter galaxies behind the Bullet Cluster} 
\authorrunning{Johansson et al.}

\begin{document}
\maketitle

\section{Introduction}
\label{sec:introduction}

\begin{figure*}[t]
  \centering 
  \includegraphics{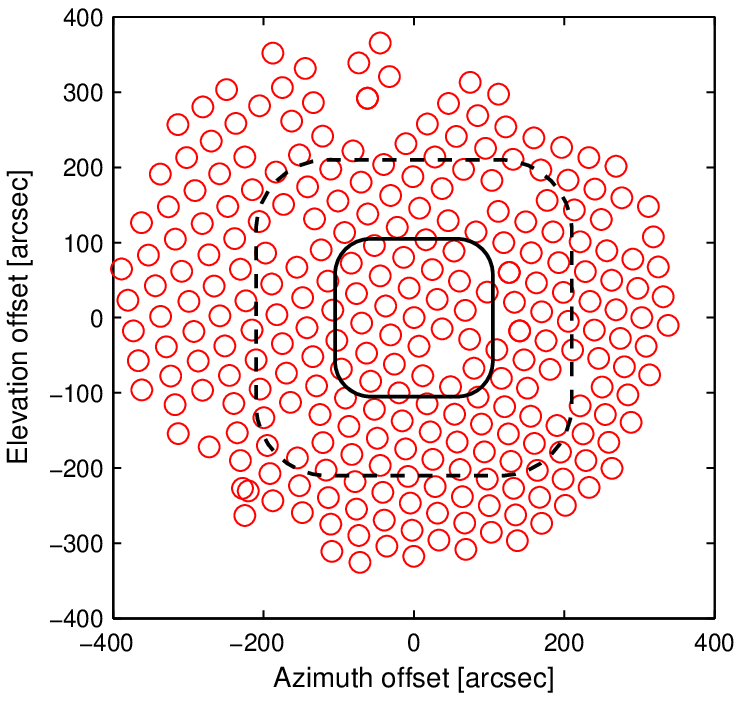} 
  \includegraphics{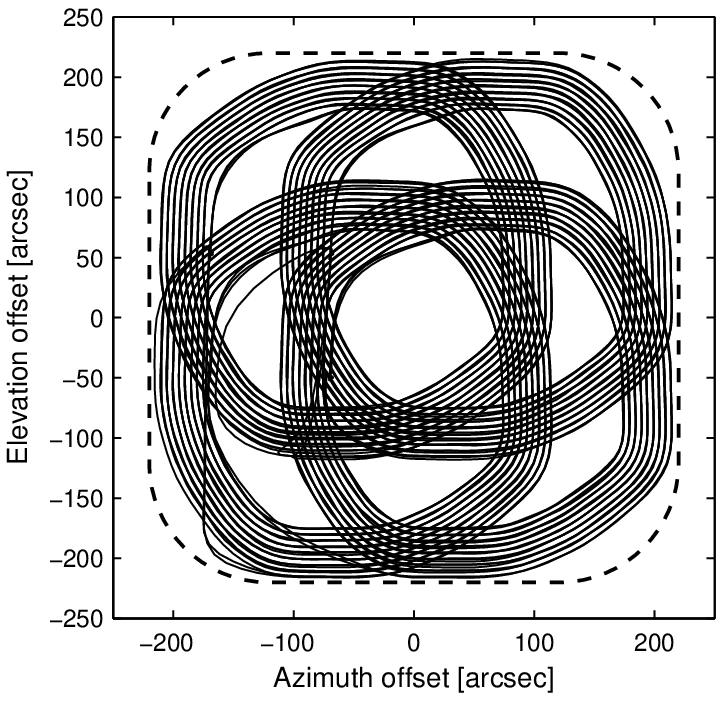}
  \caption{\emph{Left}: layout of the bolometers on the
    LABOCA~array. Each circle represents one bolometer that was active
    at the time of the observations. The two rectangles illustrate the
    sizes of the two raster+spiral scanning pattern for one bolometer
    (solid line--small pattern, dashed line--large
    pattern). \emph{Right}: movement pattern for one bolometer during
    the large scanning pattern. This pattern has a $2\times 2$ raster
    setup with 100\arcsec~between the raster points (the patterns are
    described in detail in Sect.~\ref{sec:submillimeter-laboca}). The
    smaller scanning pattern is similar, but because the distance
    between raster points is considerably smaller, it is harder to
    illustrate. A comparison between the two patterns can instead be
    made from the two rectangles in the left panel. For clarity, we
    plot also the dashed rounded rectangle of the left panel in the
    righ panel.}
  \label{fig:orbits}
\end{figure*}

The large concentrations of mass (up to $10^{15} M_\odot$) on angular
scales of a few arcminutes in galaxy clusters act as natural
gravitational lenses capable of magnifying background galaxies that
would be too dim to be detectable otherwise. In the mm and submm
wavebands, lensing by galaxy clusters makes it possible to probe the
fainter part of the brightness distribution of the so-called
submillimeter galaxies (SMGs), which are believed to be dusty
high-redshift star-forming galaxies
\citep{1997MNRAS.290..553B}. 

Pioneering observations of SMGs at 450 and $850~\mu$m were done using
SCUBA on the James Clerk Maxwell Telescope. The first observations
toward two massive clusters at $z \simeq 0.35$ resulted in the
detection of a total of six sources above the noise level of 2
mJy/beam at $850~\mu$m
\citep{SmailIvison:1997aa}. 
The authors estimated the surface density of the sources to be three
orders of magnitude larger than the expectation from a non-evolving
model using the local IRAS $60~\mu$m luminosity function.  Those
observations provided evidence for the presence of a large number of
actively star-forming galaxies at high redshift, which might be the
counterparts of the luminous and ultraluminous infrared galaxies
observed in the local universe \cite[e.g.][]{SandersMirabel:1996aa}.

At redshifts beyond one, the flux density of a redshifted
infrared-luminous galaxy is largely redshift-independent~: its
decrease with an increasing distance is compensated by the steep rise
in the mm and submm due to the redshifted spectral energy distribution
(\citealt{1993MNRAS.264..509B}). 
During the last decade, several hundreds of SMGs have been discovered
using bolometer arrays, mostly SCUBA \cite[e.g.][]{Blain:1998aa,
  BorysChapman:2003aa,
  CoppinChapin:2006aa}, 
and more recently MAMBO at 1.2~mm and AzTEC~at 1.1~mm
\citep{BertoldiCarilli:2007aa,
  ScottAustermann:2008aa,AustermannDunlop:2010aa}. The mm/submm galaxy
population is the subject of many multi-wavelength studies (see the
review by \citealt{BlainSmail:2002aa}). The median redshift of SMGs
with known redshifts is around $2-3$ \citep{SmailIvison:2000aa}.

So far, only a handful of massive galaxy clusters have been mapped in
the submm, and most of them are clusters in the northern hemisphere
observed with SCUBA. Observation of seven massive clusters with a
sensitivity of 2 mJy/beam provided a catalogue of 17 submm sources
brighter than the 50\% completeness limit
\citep{1998ApJ...507L..21S}. 
Nine other cluster fields in the redshift range 0.2 -- 0.8 were
observed with a similar sensitivity, resulting in the detection of 17
new submm sources \citep{2002MNRAS.330...92C}. 
Deeper observations with a 3-sigma limit of 1.5--2 mJy/beam of three
massive clusters probed the sub-mJy number counts, because of the
gravitational magnification of the clusters.
\citep{CowieBarger:2002aa}. 
\cite{Knudsenvan-der-Werf:2008aa} targeted twelve clusters and the New
Technology Telescope Deep Field. They detected 59 sources (some of
them being multiple images of the same galaxy), and determined that
seven of them have sub-mJy lensing-corrected flux densities.

The LABOCA~bolometer camera on APEX has been used to survey the
870~$\mu \mathrm{m}$~emission in a protocluster at $z\simeq 2.4$
\citep{BeelenOmont:2008aa} and the Extended Chandra Deep Field South
\cite{WeisKovacs:2009ab}. 
\cite{2009A&A...506..623N} 
observed the Sunyaev-Zeldovich (SZ) increment toward the massive
cluster Abell~2163 and noted one bright point source.  This is a good
candidate for an SMG lensed by the cluster.

The Bullet Cluster (1E~0657-56) at $z\simeq 0.3$ is one of the most
massive galaxy clusters known to date
\citep[see][]{Markevitch:2002lr,SpringelFarrar:2007aa}.  A bright
millimeter source was recently discovered in the Bullet Cluster field
and identified as the lensed image of a background galaxy at a
redshift of about 2.7 (\citealt{2008MNRAS.390.1061W}, hereafter W08).
The source happens to lie close to a critical line of the lens,
causing a large flux amplification.  The observations were performed
with the AzTEC~bolometer camera on the 10-meter ASTE telescope in the
Atacama desert in Chile, which provides an angular resolution of
$30''$ at 1.1~mm wavelength.  A doubly lensed source at the same
location had been previously identified in Hubble Space Telescope
($HST$) maps and was used, together with other multiply lensed
galaxies and a large number of weakly lensed sources, to obtain a
calibrated map of the projected mass distribution of the Bullet
Cluster \citep{Bradac:2006fj}.  Recently, \cite{GonzalezClowe:2009aa}
identified a third image by analyzing maps in the optical ($HST$), and
in the near- and mid-infrared (Magellan and Spitzer).  By fitting the
Spectral Energy Distribution (SED) of a starburst galaxy to the
observations, they also inferred a redshift of about 2.7. Their
lensing model gave a magnification of 10--50 for the three images.

In this paper we present results of observations of the Bullet Cluster
field at a wavelength of 870~$\mu \mathrm{m}$~using the
LABOCA~bolometer camera. At that wavelength, the emission is a
combination of extended signal due to the Sunyaev--Zeldovich~effect by
the hot intracluster gas and of point sources, which are potential
high-redshift star-forming galaxies. Recovery of the extended SZ
signal from the LABOCA~data (the SZ increment) requires a different
data reduction and will be presented in a subsequent paper. The
SZ~decrement from the Bullet Cluster~has been mapped by
e.g. \cite{2009ApJ...701...42H} with the APEX-SZ instrument, operating
at 2~mm. This paper is organized as follows: the observations are
presented in Sect.~\ref{sec:observations} and the data reduction in
Sect.~\ref{sec:data-reduction}; the results are presented and
discussed in Sects.~4 and \ref{sec:analysis}.

Throughout the paper, we adopt the following cosmological parameters:
a Hubble constant $H_0 = 70$~km~s$^{-1}$~Mpc$^{-1}$, a matter density
parameter $\Omega_0 = 0.3$, and a dark energy density parameter
$\Omega_{\Lambda0} = 0.7$. The redshift $z=0.296$ of the Bullet
Cluster corresponds to an angular-diameter distance of 910~Mpc and a
scale of 4.41~kpc/arcsec.

\section{Observations}

\subsection{Submillimeter}
\label{sec:submillimeter-laboca}

\label{sec:observations}
The observations\footnote{Swedish program ID O-079.F-9304A \\ and ESO
  program ID E-380.A-3036A} were carried out in September, October and
November 2007 using LABOCA~(Large APEX BOlometer CAmera,
\citealt{SiringoKreysa:2009aa}) on the APEX telescope\footnote{This
  publication is based on data acquired with the Atacama Pathfinder
  EXperiment (APEX). APEX is a collaboration between the
  Max-Planck-Institut f\"ur Radioastronomie, the European Southern
  Observatory, and Onsala Space Observatory.}
\citep{Gusten:2006ak}. LABOCA~is a 295-element receiver operating at a
central frequency of 345~GHz with a bandwidth of 60~GHz. At the time
of our observations, about 250 bolometers were used. The mean
point-source sensitivity of those bolometers was
78~${\mathrm{mJy~s}^{1/2}}$. The angular resolution was $19.5\arcsec$
and the field-of-view was $11.4\arcmin$. The layout of the bolometer
array is illustrated in Fig.~\ref{fig:orbits}. A total of 25 hours of
observing time was spent, including pointing and calibration. The
weather conditions were varying, with an amount of precipitable water
vapor ranging from 0.5 to 2.0~mm.

LABOCA~uses feed horn antennas and their physical size limits the
spacing of the bolometers on the array. LABOCA~is therefore not fully
sampling the sky, so the telescope has to be moved to sample the sky
such that the resulting map meets the Nyquist criterion. The scanning
pattern does not only ``fill the gaps'' between bolometers, it also
modulates the astronomical signal into a range of spatial frequencies
which facilitates filtering of $1/f$-type noise (instrumental and sky
noise).

\subsubsection{Scanning patterns}
\label{sec:scanning-patterns}

We used two different scanning patterns, as illustrated in
Fig.~\ref{fig:orbits}.  Both are a combination of Archimedian spirals
with a duration of 35 seconds, centered on a four-point raster.

-- During the first observing session, we used a compact scanning
pattern: the four points were separated by $27\arcsec$ in azimuth and
elevation, each point marking the center of a spiral with a minimum
radius of $R_0=18\arcsec$ and winding out with a radial speed of
$\dot{R}=2.25\arcsec s^{-1}$ and an angular speed of $\dot{\phi}=90
\deg s^{-1}$.  The spirals thus ranged from $18\arcsec$ to about
$97\arcsec$ in radius, with a scanning speed between $0.5\arcmin
s^{-1}$ and $2.5\arcmin s^{-1}$.

-- During the second observing session, a larger scanning pattern was
used to facilitate the retrieval of the extended
SZ~signal.  The four raster points were separated by $100\arcsec$
in azimuth and elevation, and the spirals had a minimum radius $R_0=
120''$ and radial and angular speeds of $\dot{R} = 1.25'' s^{-1}$,
$\dot{\phi} = 90\deg s^{-1}$.  The spirals thus ranged from 120 to
$164\arcsec$ in radius, with a scanning speed between 3 and $4\arcmin
s^{-1}$.

\subsubsection{Pointing, focus and calibration}
\label{sec:pointing}
The pointing accuracy was checked by repeated observations of the
nearby source PKS~0537--441. This is a variable source; during our
observations, its mean flux density was $\sim 3$ Jy. The source was
scanned in a tight spiral and the data were reduced and made into a
map using the BoA software (see Sect.~\ref{sec:data-reduction}).  A
two-dimensional Gaussian was fitted to the pointing source, and the
telescope's pointing was updated using offsets from the fit. The
pointing was stable within  $3\arcsec$.

The focus was checked at least twice during every observing session by
observing a planet (Venus, Saturn or Mars). The subreflector was moved
in small increments in each of the three cartesian directions while
the telescope tracked the source. The optimum focus position in each
direction was determined by fitting a curve to the observed points,
and the subreflector was finally moved to the position corresponding
to the maximum of the curve. 

The absolute flux calibration of LABOCA~is supposed to be accurate to 10\%
\citep{SiringoKreysa:2009aa}. We verified this by daily
observations of Uranus.

The atmospheric attenuation was determined from continuous scans in
elevation at a fixed azimuth (``skydips''), and from radiometer
measurements \citep{SiringoKreysa:2009aa}.

\subsection{Infrared}
\label{sec:infrared-spitzer}

In Sect.~\ref{sec:infr-count-lab} we describe a search for infrared
counterparts to the detected submm sources in Spitzer maps. We now
describe the data that was used for that comparison.

Spitzer IRAC and MIPS of the Bullet Cluster field were acquired from
the Spitzer data archive. Both the IRAC and MIPS data were taken under
program ID 40593 (PI: Gonzalez). The Spitzer data overlap with most of
the the region observed with LABOCA, and out of the detected submm
sources only one source lacks Spitzer coverage.

The IRAC and MIPS maps were processed by, respectively, version 18.7
and 18.12 of the SSC pipeline. We started by visually inspecting the
resulting pbcd (post basic calibrated data) mosaics, and found that
the IRAC maps had no apparent artefacts, but that the MIPS map had
clear signs of ``dark latents'', as described in the MIPS data
handbook. We therefore reprocessed the basic calibrated data (bcd)
using MOPEX version 18.3.3 the script \texttt{flatfield.pl} to
self-calibrate the bcd data. The bcd's were then mosaiced using
\texttt{mopex.pl}. The reprocessed MIPS map shows no signs of
artefacts due to ``latents''.

Properties of the acquired Spitzer data are summarized in
table~\ref{tab:spitzerdata}. We list there the median integration time
per pixel per map, the sensitivity and the angular size of each
map. The sensitivity is estimated from the final combined mosaics,
masking out all pixels brighter than 10 times the median pixel value
in each map, and then calculating the standard deviation of the
remaining pixels. The listed sensitivity values are 3$\sigma$, and are
similar to those obtained from the Spitzer Science Center
``Sensitivity -- Performance Estimation
Tool''\footnote{\texttt{http://ssc.spitzer.caltech.edu/tools/senspet/}}.

\begin{table}[h!]
  \centering
  \caption{Properties of the Spitzer data used in this study.} 
  \begin{tabular}[h]{l c c c}
    \hline
    \hline
    Band & $t_{\mathrm{int}}$ & $3\sigma$-depth  & FOV  \\
    & (s) & ($\mu$Jy) &  (arcmin$\times$arcmin) \\
    \hline
    IRAC1 & 9100 & 0.9 & 15 $\times$ 22 \\
    IRAC2 & 9100 & 1.4 & 15 $\times$ 22 \\
    IRAC3 & 9100 & 6.6 & 15 $\times$ 22 \\
    IRAC4 & 9100 & 7.0 & 15 $\times$ 22 \\
    MIPS  & 7800 & 39.5 & 13.4 $\times$ 12.7 \\
    \hline
  \end{tabular}
  \label{tab:spitzerdata}
\end{table}

\section{Data reduction}
\label{sec:data-reduction}

In this section, we describe the steps followed to produce a fully
calibrated map from the raw data, which come in the form of
time-streams containing the voltage read-outs of each bolometer as a
function of time. We have used two data reduction softwares:
{\tt Minicrush}\footnote{{\tt Minicrush}~can be downloaded from\\
  \texttt{http://www.submm.caltech.edu/{\textasciitilde}sharc/crush/download.htm}},
written originally for the SHARC bolometer array and adapted to handle
LABOCA~data \citep{2008SPIE.7020E..45K}, and
\emph{BoA}\footnote{BoA can be downloaded from \\
  \texttt{http://www.apex-telescope.org/bolometer/laboca/boa/}},
developed in Bonn (Schuller et al., in prep).  In general, the maps
produced by both pipelines were in good agreement in terms of number
and characteristics of sources; however, since the {\tt Minicrush}~map showed a
lower level of large-scale noise, we used that software for the
analysis presented in this paper.

The data consist of a total of 185 eight-minute-long scans on the
cluster, plus pointing and calibration observations.  Each scan is
contained in a separate MBFITS-file.

First, we flagged blind bolometers to exclude them from the rest of
the data analysis. Then, we corrected for differences in sensitivity
of individual bolometers. 
That information was extracted from ``beam-maps'', which are fully
sampled maps where each bolometer has scanned a bright planet.

For each scan, the zenith opacity calculated from the radiometer and
skydip measurements was used to calibrate the data for
elevation-dependent opacity variations.  We also flagged data taken
during periods of high telescope speeds and accelerations.

\subsection{Removal of correlated noise in the time-streams}
\label{sec:sky-noise-removal}

The most critical and challenging task of the data analysis is to
extract the true astronomical signal from the measurements, which are
contaminated by noise from various sources.  One component of the
noise arises in the electronic systems, such as the readouts of the
bolometer array; its spectrum is of the $1/f$ type where $f$ is the
frequency. Even more important is the contribution of the Earth's
atmosphere, which also has a $1/f$-type spectrum, but shows both
spatial and temporal fluctuations.  In the 870~$\mu
\mathrm{m}$~atmospheric window in which LABOCA operates, typical
zenith opacities at the APEX site are of the order of 0.1 to 0.2. The
atmosphere is thus largely transparent to cosmic signals; but the
amplitude of the atmospheric signal can be as high as 10$^5$ times
that of the astronomical signal of interest. Because the atmospheric
signal (or noise) is correlated across the bolometer array, it can be
estimated and removed from the time-streams.  For point-source
observations, correlated sky noise can be partly removed by filtering
low spatial frequencies (or large angular scales), where the $1/f$
noise is most severe.

The software {\tt Minicrush}~takes the following approach to remove
correlated noise from the time-streams: a correlated noise component
is modeled as a common signal in the time-streams of the different
bolometers, scaled by a gain factor which depends on each bolometer.
A $\chi^2$ function is minimized by taking its derivative with respect
to the modeled signal.  The time-streams of all the bolometers are
considered for a certain scan.  When the fit has been performed, the
estimated correlated signal is removed from the data.  The process of
estimation and removal of the correlated signal is carried out several
times. The uncertainties on the modeled correlated signal are
estimated by calculating the changes in the estimated signal when the
$\chi^2$ has increased by 1 from its minimum value. The ideas
implemented in {\tt Minicrush}~have been described in detail in
\cite{2008SPIE.7020E..45K}.

Since we were interested in compact sources, we used the option {\it
  --deep} in {\tt Minicrush}.

\subsection{Map-making}
\label{sec:filt-basel}

\subsubsection{From time-streams to maps}

\begin{figure}[t]
  \centering
  \includegraphics[width=9cm]{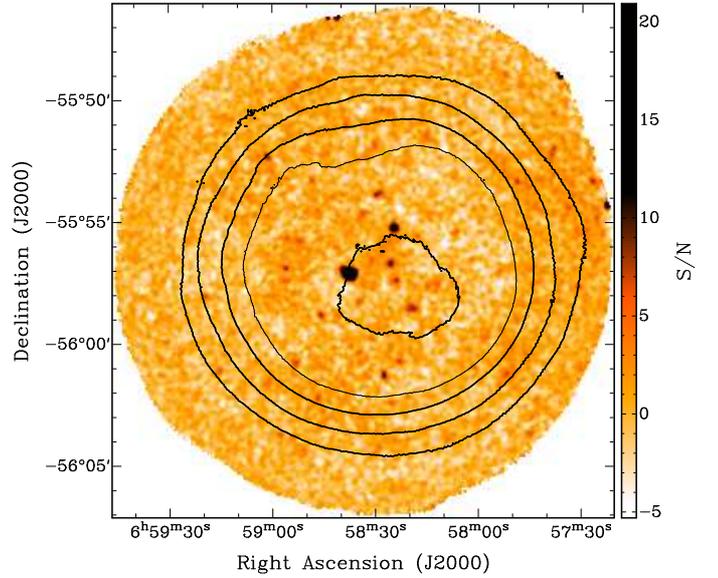}
  \caption{Color image of the 870~$\mu$m signal-to-noise map. The map
    has been filtered to remove extended signal, including that due to
    the Sunyaev-Zeldovich effect. Several sources are visible. The
    brightest coincides with the millimeter source discovered by W08
    and identified as a galaxy at redshift $z\simeq 2.7$ and strongly
    lensed by the Bullet Cluster. The contours refer to the noise
    level at 1.3, 2.1, 2.7, 4.1 and 6.8 mJy/beam. The signal-to-noise
    representation supresses the high noise levels in the outer parts
    of the map. They are instead indicated in the contours of the
    noise map.}
  \label{fig:sigweigh}
\end{figure}

The data consist of a collection of time-streams from each functioning
bolometer, cleaned from atmospheric and instrumental noise. The data
in each time-stream of a scan are used to create a map of that scan,
taking into account the scanning pattern of the telescope and using a
nearest-pixel mapping algorithm. We refer to those maps as
scan-maps. For each scan-map, a corresponding noise map is generated
by adding in quadrature the noise levels of the bolometers that hit a
certain pixel in the map. The noise maps are used to weigh the
individual scan maps when they are co-added, which is the final step
of the map-making process. The individual noise maps are also coadded
into a final noise map. This noise map reflects the pixel rms in the
data. As the noise map only accounts for the relative weights between
map pixel, we then rescaled the noise map. In
Sect.~\ref{sec:noise-properties} we show that the signal-to-noise map
has a Gaussian distribution of pixel values. We scaled the noise map
in order for the signal-to-noise map to have a pixel histogram with a
standard deviation of 1. This process scales the noise map with a
factor of 1.5. The rescaled noise map is shown as contours in
Fig.~\ref{fig:sigweigh}, together with the the signal-to-noise map
which is constructed by dividing the signal map with the scaled noise
map.

The maps have a pixel size of 4\arcsec. This means that the
oversampling factor compared with the original resolution of the
observations is $19.5/4 \sim 5$. Such a fine pixelization is
preferential for the map-making process to be effective. The resulting
map has a pixel-to-pixel noise level which is affected by high
frequency noise on a scale smaller than the beam. This noise (which is
due to small pixelization) can easily be removed by smoothing. We
smoothed the final maps using a 10\arcsec~Gaussian in order to remove
that high-frequency noise component and produce cleaner maps.

\subsubsection{Iterative mapping}

Some artefacts such as ``sidelobes" around point sources are seen in
the final map because of the filtering. In order to remove those
artefacts, the entire {\tt Minicrush}~reduction was applied a second
time, but instead of building a source model from the actual data
being reduced, we used a source model based on the results of the
first reduction.  The part of the map with a signal above $4.5\sigma$
was used as the source model.

\begin{figure}[t]
  \centering
  \includegraphics{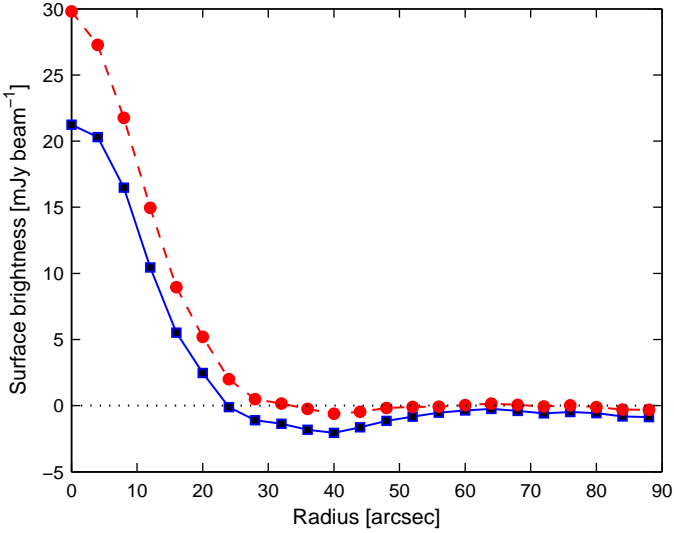}
  \caption{Radial profiles of the brightest source (source \#1): in
    the first iteration of the data reduction (solid line and boxes),
    a source model was successively built during the data reduction;
    in the second iteration (dashed line + circles), a
    $4.5\sigma$-clipped version of the first map was used as a source
    model. The iterative method makes it possible to recover flux that
    was lost at the first stage of the data reduction.}
  \label{fig:azprof}
\end{figure}

Figure~\ref{fig:azprof} illustrates the result of that iterative
process: radial profiles of the brightest source in the map are shown
for each of the two iterations. The difference between the two
profiles is significant. We have observed that the difference is
greater the brighter the source. The iterative process makes it
possible to recover the flux that was lost in the first iteration. We
have compared the fluxes of our two most significant sources (after
two iterations) with the values obtained using the standard data
reduction pipeline that was used for the calibrator, Uranus. The
agreement showed that two iterations were sufficient, as also found by
other groups \citep[e.g.][]{WeisKovacs:2009ab}.

\section{Results}

Figure~\ref{fig:sigweigh} shows the signal-to-noise map obtained from
the data reduction process described above. Several sources are
visible. Since the noise level increases steeply toward the outer
parts, sources with low signal-to-noise ratio in that part of the map
may have high flux densities. In this section, we discuss the noise
properties of the data and present the methods that we have used to
identify sources and measure their properties. We also present Monte
Carlo simulations performed to quantify the degree of completeness to
which sources can be extracted, and to estimate the amount of flux
boosting due to the confusion noise. Finally, we estimate the
magnification of each source due to gravitational lensing, using a
simple model of the Bullet Cluster and assuming that all sources are
at a redshift of 2.5.

As shown in the following analysis, neither completeness nor flux
boosting corrections change the results very much, because of the
conservative detection treshold that we have adopted. The results is a
fairly robust catalog of 17 sources.

\subsection{Noise properties}
\label{sec:noise-properties}

\begin{figure}[t]
  \centering
  \includegraphics{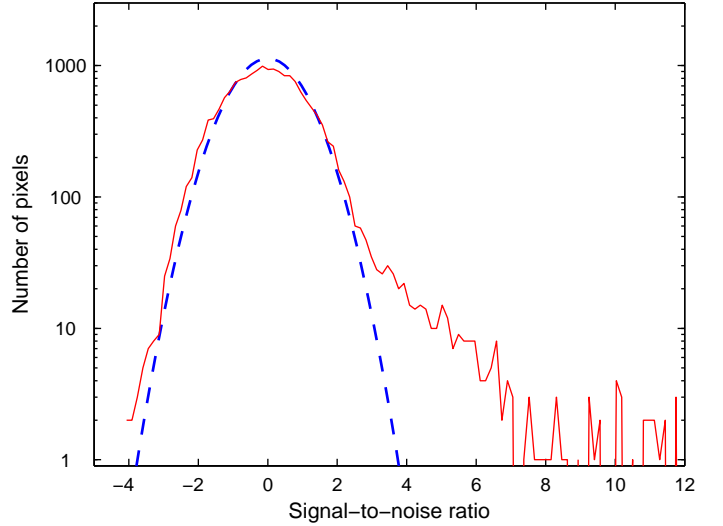} 
  \caption{ \emph{Solid curve}: Histogram of the pixel distribution in
    the central 10 arcminutes of the signal-to-noise map. \emph{Dashed
      curve}: Histogram of the mean pixel distribution in the same
    region of the jackknife maps, as described in the text. The
    jackknife procedure has removed the excess positive signal (due to
    the sources) and the negative signal (due to the sidelobes) that
    are present in the signal map.}
  \label{fig:hist}
\end{figure}

To quantify the noise level in the map, we constructed 500 so-called
jackknife noise maps, which are obtained by multiplying half of scan
maps (selected randomly) by $(-1)$, before co-adding all the scan
maps. The jackknife maps should therefore be free from astronomical
sources and from artefacts from the data reduction pipeline, and
reveal the nature of the statistical noise in the data.  Each
jackknife map was smoothed with a 10\arcsec~Gaussian after coadding,
giving a final resolution of $\sqrt{(19.5\arcsec)^2+(10\arcsec)^2}
\simeq 22$\arcsec. Then, we calculated the mean of the 500 histograms
of the pixel distributions measured in the central 10\arcmin~of each
jackknife map.

Figure~\ref{fig:hist} shows the mean histogram, in units of
signal-to-noise value, together with the histogram of the pixel
distribution of the signal-to-noise map, extracted from the same
central 10\arcmin. The jackknife histogram is fitted by a Gaussian
with standard deviation 1 (as expected). If we instead make the same
calculation in the signal map, we find that the jackknife histogram is
well fitted by a Gaussian function with mean $\mu = 1.4\times
10^{-6}$~mJy and standard deviation $\sigma = 1.17$~mJy. Therefore,
when excluding astronomical sources and systematic effects from the
data reduction, the statistical noise level is 1.2 mJy/beam.

\begin{table*}[t]
  \centering
  \caption{List of sources extracted from the LABOCA map.} 
  \begin{tabular}[t]{l c c c c c c c }
    \hline \hline
    Source & $\alpha$(J2000) & $\delta$(J2000) & Flux
    density\tablefootmark{a} & Deboosted\tablefootmark{b} &
    Demagnified\tablefootmark{c} & $F/\Delta F$\tablefootmark{d} \\
		& & & & flux density &flux density \\
    & (h:m:s) & ($^\circ:':''$)  & (mJy) & (mJy) & (mJy) & \\
    \hline
    1 & 06:58:37.62 & -55:57:04.8 & $48.6\pm 1.3$\tablefootmark{e} & $ 48.0\pm 1.3$ & 0.64 & 82.8\\
    2 & 06:58:24.47 & -55:55:12.5 & $15.1\pm 1.0$ & $ 14.7\pm 1.0$ & 8.8 & 29.9\\
    3 & 06:58:25.45 & -55:56:40.1 & $6.9\pm 0.9$ & $ 6.4\pm 1.0$ & 2.2 & 17.8\\
    4 & 06:58:19.36 & -55:58:30.3 & $8.2\pm 0.9$ & $ 7.7\pm 0.9$ & 4.7 & 16.2\\
    5 & 06:58:27.27 & -56:01:16.3 & $9.0\pm 1.3$ & $ 8.0\pm 1.3$ & 6.3 & 15.9\\
    6 & 06:58:28.94 & -55:53:48.4 & $9.3\pm 1.2$ & $ 8.6\pm 1.2$ & 6.3 & 15.4\\
    7 & 06:59:01.39 & -55:52:18.1 & $11.9\pm 2.1$ & $ 9.7\pm 2.1$ & 8.4 & 14.2\\
    8 & 06:58:24.05 & -55:57:23.0 & $5.3\pm 0.9$ & $ 4.7\pm 1.0$ & 1.8 & 13.1\\
    9 & 06:58:55.98 & -55:56:51.7 & $5.4\pm 1.2$ & $ 4.4\pm 1.3$ & 3.3 & 12.8\\
    10 & 06:58:45.60 & -55:58:48.0 & $6.2\pm 1.1$ & $ 5.5\pm 1.1$ & 3.6 & 12.0\\
    11 & 06:58:53.22 & -56:00:45.0 & $7.8\pm 1.5$ & $ 6.4\pm 1.6$ & 5.2 & 11.9\\
    12 & 06:58:52.22 & -55:55:45.7 & $5.5\pm 1.2$ & $ 4.5\pm 1.2$ & 3.4 & 11.2\\
    13 & 06:58:22.88 & -56:00:40.7 & $4.8\pm 1.2$ & $ 3.8\pm 1.3$ & 2.9 & 11.0\\
    14 & 06:58:46.68 & -56:02:11.8 & $7.2\pm 1.9$ & $ 4.6\pm 2.5$ & 3.8 & 10.8\\
    15 & 06:58:33.69 & -55:54:40.8 & $4.6\pm 1.1$ & $ 3.6\pm 1.2$ & 2.5 & 10.1\\
    16 & 06:58:12.44 & -55:57:29.7 & $4.9\pm 1.0$ & $ 4.2\pm 1.0$ & 1.9 & 9.2\\
    17 & 06:59:15.72 & -56:01:07.5 & $23.6\pm 5.9$ & ---\tablefootmark{f} & --- & 9.0\\
    \hline
  \end{tabular}
  \tablefoot{Statistical uncertainties on the
    listed positions are 1--2\arcsec, which is smaller than the pointing
    uncertainty. \tablefoottext{a} Flux density as extracted from the
    map. \tablefoottext{b} Flux density corrected for boosting due to
    confusion noise. \tablefoottext{c}Flux density corrected for
    lensing. \tablefoottext{d} Significance of the detection in the
    Gaussian-matched-filtered map. \tablefoottext{e} Source~\#1 is
    extended relative to the $22''$ beam: it has an apparent size of $29.2''\times
    23.3''$. \tablefoottext{f} Source~\#17 lies in the outer part of the map where the noise level is high 
    and the method used to deboost the flux densities fails.  }
  \label{tab:sources}
\end{table*}

\subsection{Source extraction}
\label{sec:source-extraction}

Although the noise in our final map is fairly uniform across the
central 10 arcminutes, it increases slightly with radius across
that area, and rapidly outside (see the noise map contours in
Fig.~3). In order to identify significant sources in the map, a
well-defined and mathematically justified algorithm must be used.
Following other authors \cite[e.g.][]{BeelenOmont:2008aa}, we used the
so-called ``Gaussian matched filter'' (GMF) technique outlined by
\cite{SerjeantDunlop:2003aa}.  This method is optimal for point source
extraction in a $\chi^2$ sense, although the performance is degraded
for crowded maps \citep{SerjeantDunlop:2003aa}.

The GMF significance map, $F/\Delta F$, is computed as
\begin{equation}
  \label{eq:1}
  \frac{F}{\Delta F} = \frac{(S \cdot W) \otimes P}{\sqrt{W \otimes P^2}} 
\end{equation}
where $S$ is the signal map, $W$ is the weight map (the reciprocal of
the noise map, squared), and $P$ is a Gaussian of the same size as the
beam. The $\otimes$ sign denotes a convolution.

We generated two GMF maps, one for each iteration performed in
{\tt Minicrush}~(see Sect.~3.2.2).  Setting a threshold $F/\Delta F > 9$, which
roughly corresponds to a signal-to-noise level of 4, we extracted 19
sources in the first GMF map and 22 sources in the second.  We decided
to be conservative and not include in the final catalog the sources
that had appeared in the GMF map of the iterated map.  In addition, we
excluded the two sources that were present in the first GMF map but
not in the second. Our final source list thus contains 17 sources.

\begin{figure}[t!]
  \centering
  \includegraphics[width=8.8cm]{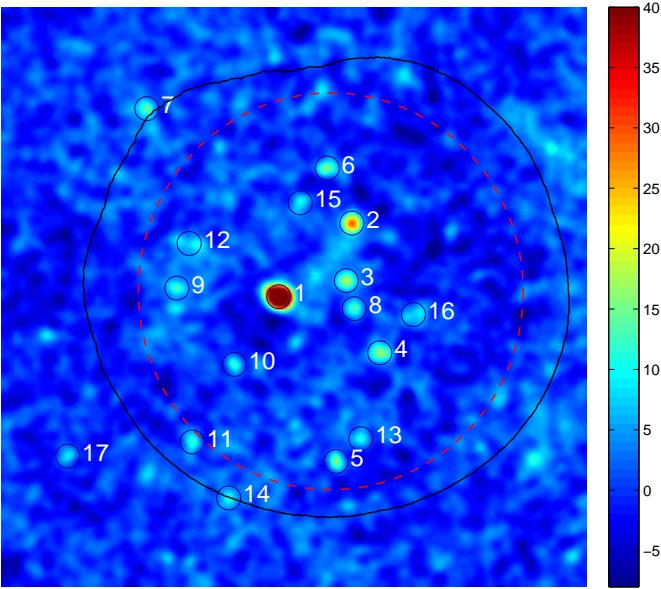}
  \caption{Gaussian filtered map, with the 17 detected sources marked
    with circles. The numbering of the sources is the same as in
    Table~\ref{tab:sources}. The black contour corresponds to the 2
    mJy/beam level in the noise map and the dashed circle marks the
    central 10\arcmin~area of constant noise level used in the
    analysis. }
  \label{fig:gf}
\end{figure}

Figure~\ref{fig:gf} shows the GMF map calculated using the iterated
signal map. The black contour indicates the 3~mJy/beam noise level,
and the red circle has a diameter of 10~arcminutes. The 17 identified
sources are marked.

Whereas source finding was done in the GMF map, measurement of the
properties of the identified sources was done in the signal map.  The
following scheme was used:

\begin{enumerate}
\item We searched for the peak value in the GMF map and fitted a
  two-dimensional Gaussian.
\item The position obtained from that fit was used as a starting guess
  in the fit to the real source in the signal map. The fitted
  function is the sum of a two-dimensional Gaussian and a tilted
  plane. The use of a plane reduces the effect of remaining sidelobes
  around bright sources. Those are due to the filtering in the data
  reduction and are not completely removed by the iterative method
  described in Sect.~3.2.2 (Fig.~\ref{fig:azprof}). Unless the source
  is clearly extended, the width of the Gaussian was fixed to that of
  the beam.
\item The integrated flux density of each extracted source was
  calculated by integrating only the Gaussian part of the fitted
  function.
\item The fitted Gaussian was then subtracted from the GMF map and a
  new search for the peak was performed.
\item We iterated over the previous steps until the peak value in the
  GMF map was smaller than 9. This value was inferred from the
  simulations described in the next section.
\end{enumerate}

Table~\ref{tab:sources} lists the properties of the 17 sources, sorted
by decreasing value of $F/\Delta F$, given in the last column.  The
fitted positions and integrated flux densities are also given.  All
sources but Source~\#1 are point sources.  The uncertainties on the
extracted flux densities are estimated from the 500
jackknife maps.

\subsection{Completeness}
\label{sec:completeness}

In order to understand the systematics of the source extraction
procedure and quantify the extent to which the values of $\Delta F/F$
correspond to real sources, we turned to simulations. A Gaussian
source of the size of the beam was added at a random location within
the central 10\arcmin-diameter region of a randomly selected jackknife
map, and the Gaussian filtered map was produced, using Eq.
(\ref{eq:1}). We stepped through a range of values of the flux
densities of the simulated source, ranging from 1 to 15~mJy, with a
spacing of 0.5~mJy. For each flux density value, 500 sources were
simulated and placed at a random location, and the GMF maps were used
to find the simulated sources. Once they were found, a two-dimensional
circular Gaussian was fitted to the simulated signal map at the same
location. For each flux density bin, we then extracted information
about the completeness and the recovered flux as a function of the
input flux density value.

\begin{figure}[t]
  \centering
  \includegraphics{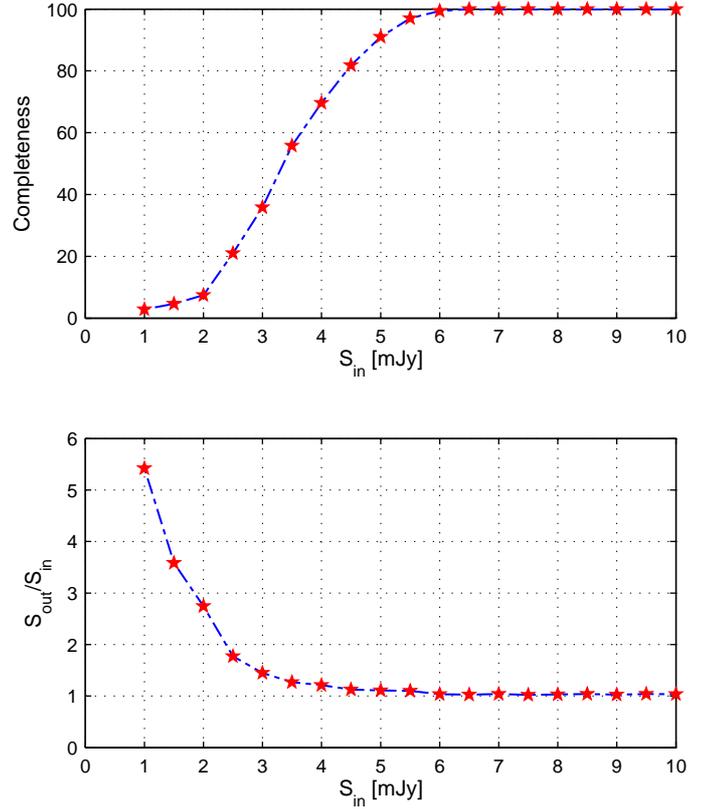}
  \caption{Results of the Monte Carlo simulations performed to
    estimate the degree of completeness of the source extraction
    algorithm and the accuracy of the measured flux density at a given
    value of the input flux density. }
  \label{fig:sim}
\end{figure}

Figure~\ref{fig:sim} shows the results. The flux boosting is
calculated as the fraction of the measured flux of a source to the
input flux density. At flux densities larger than 6 mJy we see that,
in the absence of confusion noise, no boosting is observed. For lower
flux densities the flux boosting becomes larger than one, so that
extracted sources are those that are placed on positive noise peaks,
making them rise above the noise. Sources randomly placed in noise
voids are not extracted.

The completeness is defined as the fraction of sources that are
recovered from the simulated maps out of the 500 input sources. We see
that at 6 mJy the observations are 100\% complete. At the level of 4.6
mJy, which is the flux of the most dim source in our sample, the
completeness is $\sim 85\%$. Therefore, our sample is highly complete,
and completeness corrections are not very important.

Other groups use instead the final signal map when simulating the
completeness (eg. \citealt{ScottAustermann:2008aa}), but we wanted to
focus our attention in these simulation on the effect of statistical
noise in the map. No source confusion noise is present in the
jackknife maps. In the following section, we discuss the effect of
confusion noise and use noise-free sky realizations with a Schechter
distribution to quantify the flux boosting for an individual source
extracted from the real map.

\subsection{Flux boosting due to confusion noise}
\label{sec:flux-boosting-due}

Several SMG surveys have shown that the number counts steepen towards
higher flux densities (e.g.
\citealt{2006MNRAS.370.1057S,CoppinChapin:2006aa}). 
Therefore there are many more sources at low than at high flux
densities. Most of those faint sources are below the noise level of
submm maps, but they influence photometric measurements of extracted
sources, acting as a ``sea'' of sources, often referred to as
``confusion noise''. This effect has been discussed by e.g.
\cite{Condon:1974aa} and \cite{HoggTurner:1998aa}.  Recently,
\cite{CoppinHalpern:2005aa} discussed the effect of confusion noise on
flux boosting in the SCUBA Groth strip survey, and used a Bayesian
technique to ``deboost'' the fluxes. We employ a similar technique to
estimate the amount of flux boosting for our detected sources, which
we describe in Appendix~\ref{sec:flux-deboosting}. The derived flux
boosting corrections are small for most sources. The deboosted flux
densities are listed in Table~\ref{tab:sources}.

\subsection{Lensing correction}
\label{sec:lensing-correction}

\begin{figure*}[t]
  \includegraphics[height=7.2cm]{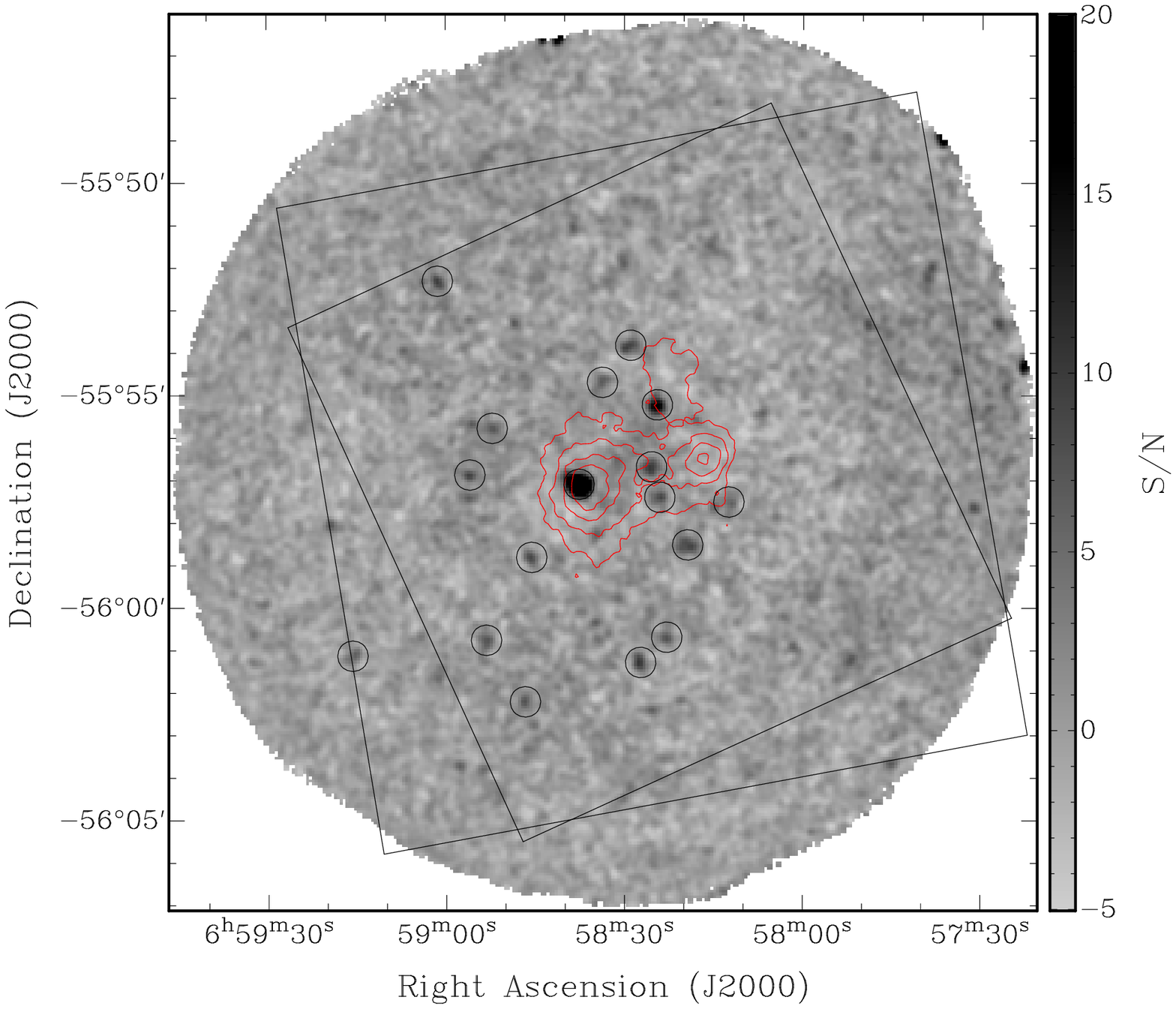}
  \includegraphics[height=7.2cm]{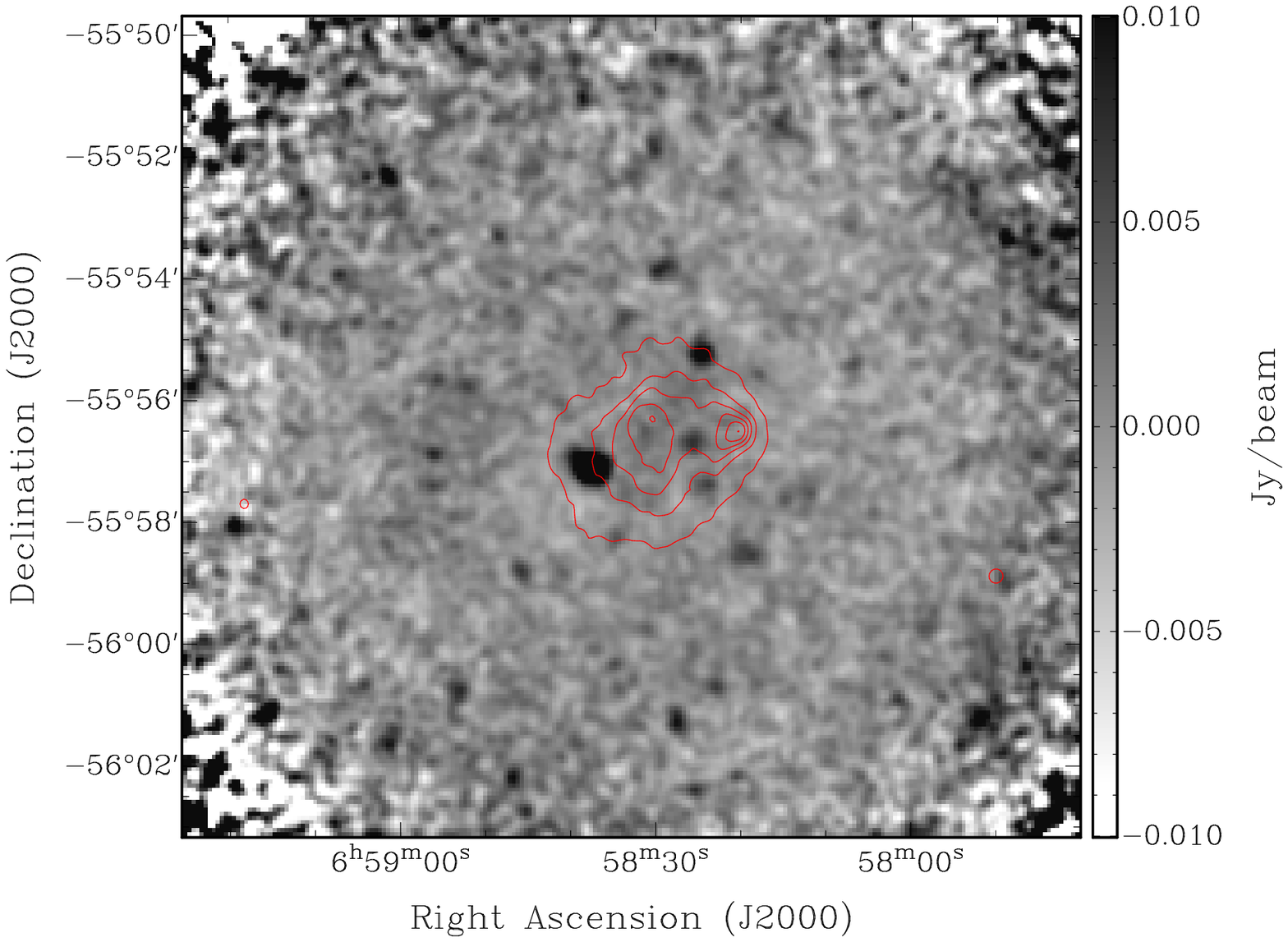}

  \caption{\emph{Left}: Signal-to-noise map with overlaid circles
    indicating extracted submm sources. The contours show the the
    projected mass density from the weak lensing analysis by
    \cite{Clowe:2006th}. The contours range from 40 to 85\% of the
    maximum value and are spaced by 15\%. The weak lensing map was
    retrieved from the website
    \texttt{http://flamingos.astro.ufl.edu/1e0657/public.html}. The
    rectangles show the regions of complete coverage of Spitzer MIPS
    (small rectangle) and IRAC (large rectangle). \emph{Right}: Signal
    map (in units of Jy/beam) with contours of the X-ray surface
    brightness from XMM-Newton observations. The noise level in the
    signal map increases rapidly towards the outskirts because of the
    low coverage there. }
     \label{fig:wlxray}
\end{figure*}

We built a simple lensing model of the Bullet Cluster in order to
estimate the magnification of our observed submm sources.  The model
consists of two spherically symmetric components representing the main
component of the Bullet Cluster and the subcomponent (the actual
``bullet'' to the west).  Figure~\ref{fig:wlxray} (left) shows contours
of the projected mass distribution inferred from weak-lensing
observations by \cite{Clowe:2006th}, overlaid on our 870~$\mu$m image.
We used the fits to the weak-lensing observations made by
\cite{Bradac:2006fj} 
to the masses within a certain projected radius $R$ of each component,
$M(< R)= M_{0} (R/250~{\rm kpc})^{n}$, with $M_{0} = 0.22\times
10^{15} M_\odot$ and $n=0.8$ for the main cluster and $M_{0} =
0.17\times 10^{15} M_\odot$ and $n= 1.1$ for the subcluster. To place
the two mass components, we used the information given in Table~2 of
the paper by \citeauthor{Bradac:2006fj}  
The redshift of our simulated Bullet cluster was set to 0.296 and that
of the source plane to $z=2.5$.  Note that the magnification values
are not very sensitive to the redshift of the sources: varying it from
$z=2$ to $z=3$ changes the magnifications by less than 10\%.  The
magnification map was calculated following the derivation in the book
by \cite{1992grle.book.....S}. 
This numerical calculation provides results along the lines of those
obtained analytically for two point masses by
\cite{1986A&A...164..237S} 
and for two isothermal spheres by
\cite{2008MNRAS.390..505S}, 
but in the case of two power-law projected mass distributions.
Because the model does not include lensing by the individual cluster
galaxies, the location of the critical lines differ from the observed
ones by about 10~arcsec.  The true magnification for a given source
must therefore differ from our derived values.  Nevertheless, this
simple model makes it possible to estimate the individual
magnifications and the average magnification in a certain
region. Using the calculated magnifications, we corrected the measured
flux densities. The lensing-corrected values are listed in
Table~\ref{tab:sources}.

\begin{figure*}[h]
  \centering
  \includegraphics[width=18cm]{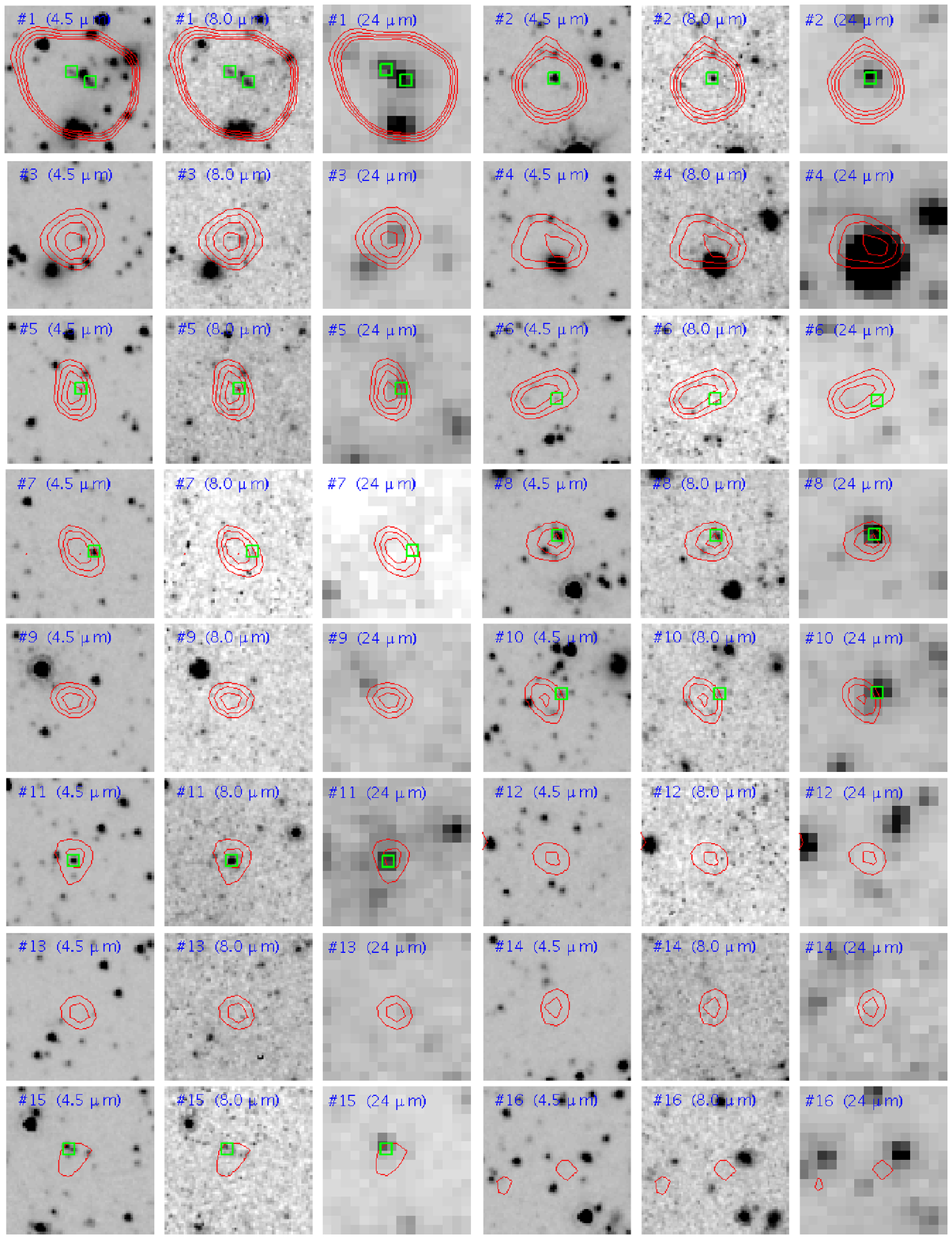}
  \caption{ 1\arcmin$\times$1\arcmin cutouts of the Spitzer 4.5, 8.0
    and 24~$\mu$m maps centered on the positions of the
    LABOCA~sources.  The contours correspond to the 3.5, 4.5, 5.5 and
    $6.5\sigma$-levels in the LABOCA~map. The grey-scales for the
    Spitzer images are spaced between $[0.0, 0.4]$, $[2.4,2.8]$ and
    $[20.25, 20.65]$ MJy/sr~respectively and the pixel sizes are 0.85,
    0.85 and 3.5 arcseconds. We have also marked the positions of the
    likely infrared counterparts to the submm sources. The counterpart
    association is described in detail in
    Sect.~\ref{sec:infr-count-lab} and~\ref{sec:notes-indiv-sourc}.}
  \label{fig:post}
\end{figure*}

\section{Discussion}
\label{sec:analysis}

\subsection{Searching for infrared counterparts to the submm sources}
\label{sec:infr-count-lab}

The identification of optical counterparts to SMGs is often difficult
due to their optical faintness, as much of the starburst luminosity is
highly obscured by dusty clouds.

Rest-frame infrared photometry of SMGs is usually possible because the
extinction is much smaller than in the optical. With the Spitzer
satellite and its imaging photometers IRAC \citep{FazioHora:2004aa}
and MIPS \citep{RiekeYoung:2004aa}, it is now possible to obtain
high-resolution infrared images of these distant objects. The number
counts of sources at 3.6~$\mu \mathrm{m}$~is high and counterpart
identification is complicated by the large positional uncertainty of
the submm source (see below). In the 24~$\mu \mathrm{m}$~band, the number
density of sources is smaller and a more secure counterpart
identification can be made, if a source can be identified.

Figure~\ref{fig:post} shows cutouts of IRAC~4.5 and IRAC~8.0 and MIPS
24.0~$\mu \mathrm{m}$~images of 16 sources detected in the
LABOCA~map. Source~\#17 has only partial Spitzer coverage and is omitted
from this figure. The Spitzer data used in this study are described in
Sect.~\ref{sec:infrared-spitzer} and the coverage with respect to the
LABOCA~map is shown in Fig.~\ref{fig:wlxray}.

The full 1\arcmin$\times$1\arcmin cutouts are only shown for reference
here. It is not necessary to search for infrared counterparts in such
a large field. The region of interest (within a certain search radius)
is the sum of the contributions from the mean pointing error of
LABOCA~of $\sim 4$\arcsec; the statistical uncertainties on the fitted
positions, which are $~1-2$\arcsec; the systematic uncertainty due to
confusion noise, $\sim 3-4$\arcsec~and a possible misalignment between
the LABOCA~and Spitzer maps of the order of $~5$\arcsec. Together, this
adds up to a search radius of $\sim 10$\arcsec.

Table~\ref{tab:spitzer} lists the coordinates of likely Spitzer
counterparts, found using the software SExtractor
\citep{1996A&AS..117..393B}, 
and their measured flux densities in the 3.6, 4.5, 5.8, 8.0, and
24~$\mu \mathrm{m}$~bands. We extracted sources with six adjacent
pixels above $3.5\sigma$, and used an aperture of five pixels for the
photometry; applying the aperture corrections listed in the IRAC and
MIPS data handbooks. The listed uncertainties on the flux densities
are the statistical errors given by SExtractor. We estimate the
systematic uncertainties to be $\sim 10\%$. In the cases where a
source has been extracted in one IRAC channel and not another we list
upper limits.

We found in total 9 sources with infrared counterparts, and two where
counterpart identification is complicated by the larger number of
sources in the short wavelenght IRAC bands. For two sources we see an
excess of flux in the MIPS map, but it is not significant enough to be
extracted. Deeper MIPS imaging would therefore be useful. Deep
high-resolution radio maps could also be used to distinguish between
sources in the shorter wavelength bands. Comments about the individual
sources are presented below. The lack of infrared counterparts was to
be expected: for comparison, in the SHADES survey
\cite{PopeScott:2006aa} found 21 secure Spitzer counterparts to the 35
submm sources. An important result is that the large positional
uncertainty of LABOCA~($\sim10$\arcsec~for most sources) is reduced
because the pointing accuracy of Spitzer is less than $1''$.

Out of the nine sources with likely infrared counterparts, five have
sufficient coverage to make it possible to investigate the shape of
the mid-infrared SED. \cite{IvisonGreve:2004aa} suggested a diagram
based on mid-infrared colors $S_{8.0}/S_{4.5}$ versus
$S_{24}/S_{8.0}$. Based on the redshift tracks (the position as a
function of redshift for a SED in color-color space) of typical
starburst and AGN-type spectral energy distributions, such a diagram
could distinguish between strong starburst SEDs and powerful AGN. A
similar diagram was also used by \cite{IvisonGreve:2007aa} and
\cite{BeelenOmont:2008aa}, while \cite{HainlineBlain:2009aa} showed
that its diagnostic capacity is limited. The five sources with
identified counterparts in these three Spitzer bands have colors that
lie in the starburst part of the diagram. This does not exclude the
possibility of contributions from AGN to the dust heating, but
indicates that the galaxies are starburst, rather than AGN,
dominated. Infrared spectroscopic measurements could be used to
investigate further the power source of those galaxies.

\begin{table*}[t]
  \centering
  \caption{Photometry of possible infrared counterparts to the LABOCA sources.}
  \begin{tabular}[t]{l c c c c c c c c}
    \hline \hline
    Source & $\alpha(J2000)$ & $\delta(J2000)$ &  $d_{\mathrm{LABOCA}}$\tablefootmark{a} & $S_{\mathrm{3.6
        \,\mu m}}$ & $S_{\mathrm{4.5 \, \mu m}}$ & $S_{\mathrm{5.8 \, \mu m}}$  & $S_{\mathrm{8.0 \, \mu m}}$ & $S_{\mathrm{24 \, \mu m}}$ \\
    & (h:m:s)  & ($^\circ:':''$) & (\arcsec) & ($\mu$Jy) &  ($\mu$Jy) &
    ($\mu$Jy) & ($\mu$Jy)  & ($\mu$Jy) \\
    \hline
    2 & 06:58:24.59 & -55:55:12.2 &   1.1 & $ 71.2 \pm   0.3$ & $ 99.5
    \pm 0.5$ & $ 77.8 \pm   1.6$ & $ 53.1 \pm   1.9$ & $507.5 \pm  20.5$  \\ 
    5 & 06:58:27.17 & -56:01:16.8 &   1.0 & $ 12.1 \pm   0.2$ & $ 15.1
    \pm   0.3$ & < 19.9  & $ 24.9 \pm   1.5$ & $249.6 \pm  24.3$  \\ 
    6 & 06:58:28.92 & -55:53:52.0 &   3.6 & $ 13.7 \pm   0.4$ & $  8.8
    \pm   0.5$ & < 6.0 & < 2.0 & < 40 \\ 
    7 & 06:59:00.72 & -55:52:21.9 &   6.8 & $ 89.2 \pm   0.2$ & $ 54.6
    \pm   0.4$ & $ 26.8 \pm   1.5$ & < 18.7 & < 40 \\ 
    8 & 06:58:23.96 & -55:57:19.5 &   3.6 & $ 55.7 \pm   0.6$ & $ 68.5 \pm   0.6$ & $ 50.1 \pm   1.5$ & $ 40.5 \pm   1.5$ & $582.3 \pm  24.4$  \\ 
    10 & 06:58:45.39 & -55:58:46.4 &   2.3 & $ 17.5 \pm   0.7$ & $
    23.7 \pm   0.7$ & $ 28.2 \pm   1.9$ & < 18.4 & $608.0 \pm  39.0$ \\ 
    11 & 06:58:53.52 & -56:00:48.4 &   4.3 & $ 33.4 \pm   0.4$ & $ 47.0 \pm   0.4$ & $ 72.2 \pm   2.2$ & $ 88.5 \pm   1.6$ & $228.8
    \pm  33.5$ \\ 
    15 & 06:58:34.14 & -55:54:37.2 &   5.3 & $ 46.1 \pm   0.3$ & $
    37.2 \pm   0.4$ & $ 37.4 \pm   1.4$ & $ 30.1 \pm   2.1$ & $301.3 \pm  15.5$  \\ 
    \hline
  \end{tabular}
  \tablefoot{\tablefoottext{a} Distance between the IRAC 1 position
    and the central LABOCA position.}
  \label{tab:spitzer}
\end{table*}

\subsection{Notes on individual sources}
\label{sec:notes-indiv-sourc}

Aside from Source~\#1, none of the other sources have been detected
previously in the mm or submm.  Our observation of Source~\#1 is
discussed in the context of other observations.  The few other sources
for which complementary observations exist are discussed as well.

\textbf{Source \#1:} With a deboosted flux density of
$48.0\pm1.3$~mJy, this source is one of the brightest SMGs ever
detected around 870~$\mu \mathrm{m}$. This is very likely because of its
proximity to a caustic line, which provides a large magnification.
From their lensing model, \cite{GonzalezClowe:2009aa} estimated that
the two brighter images of the galaxy, A and B, which are separated by
$8.6''$, have a magnification of 25 and 50, whereas the third image,
C, located $47''$ away from the first one, would have a magnification
four times lower than the first one. Our final map was smoothed to
$22''$ and our fit to the Source~\#1 gives a position between the
images A and B; the flux that we measure comes most likely from those
two images. The total magnification is therefore 75.

In the LABOCA~study of the protocluster J2142-4423 by
\cite{BeelenOmont:2008aa} the brightest source has a flux density of
21.1~mJy.  The flux density at 850~$\mu \mathrm{m}$, which is one of
the SCUBA wavelengths, can be extrapolated using a submm spectral
index of 2.7, giving a flux density 7\% higher than at 870~$\mu
\mathrm{m}$. The SCUBA flux of Source~\#1 would therefore be around
51~mJy. For comparison, the brightest source in the SHADES survey
\citep{CoppinChapin:2006aa} has 22~mJy (and a signal-to-noise ratio of
4.9).  In the submm survey of massive clusters of galaxies,
\cite{Knudsenvan-der-Werf:2008aa} detected two bright sources towards
Abell~478 and Abell~2204, with flux densities of 25.0 and 22.2~mJy
respectively.

Source \#1 was detected using the AzTEC~bolometer array at 1.1 mm
(W08); we can compare the astrometric position, angular size and
integrated flux density with the values measured at 870~$\mu
\mathrm{m}$.

\begin{itemize}
\item \textbf{Position}: the distance between the estimated central
  position of the two sources is 4.1\arcsec. This is well within the
  pointing error margins of the two telescopes.
\item \textbf{Source size}: W08 reports a source size of $36\pm 1.3
  \arcsec \times 32 \pm 1.2\arcsec$. To compare that with the
  LABOCA~source, we smoothed our map to the AzTEC~resolution of
  30\arcsec, and fitted a two-dimensional Gaussian. The fitted FWHMs at
  this resolution are $36.2 \pm 1.3\arcsec \times 30.1 \pm 1.2
  \arcsec$, in good agreement with the value of W08.
\item \textbf{Flux density}: After removal of the contribution from
  the Sunyaev-Zeldovich effect, the flux density of the source at
  1.1~mm was estimated to 13.5~mJy \citep{2008MNRAS.390.1061W}. This
  value, however, is too low and is being revised to about 20~mJy
  (G. Wilson, priv. comm.), which would give a
  spectral index $\alpha_{1100}^{870} = 3.7$ between the two
  measurements.
\end{itemize}

Far-infrared observations of this source by the BLAST experiment were
recently published by \cite{RexAde:2009aa}. BLAST observed at
wavelengths 250, 350 and 500~$\mu \mathrm{m}$, with FWHM beam sizes of
36, 42 and 60\arcsec, respectively. High-significance detections were
reported in all three bands; the flux measurements were contaminated
by that of an elliptical galaxy, which is a spectroscopically
confirmed member of the Bullet Cluster and which had to be modeled and
subtracted. After making correction for the SZ flux and color (due to
the width of the BLAST filters), the flux densities in the three bands
are $94\pm 30$, $96\pm 27$ and $110\pm 21$ mJy for bands ranging from
shorter to longer wavelengths.

\begin{figure}[t]
  \centering
  \includegraphics{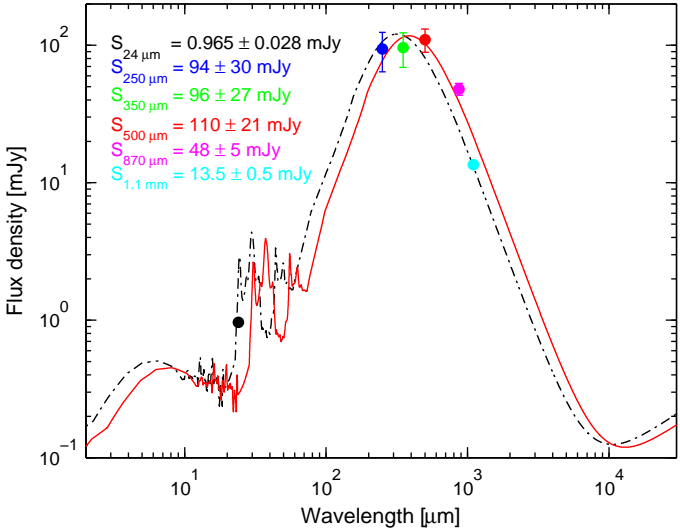}
  \caption{Spectral energy distribution of Source~\#1.  The curves
    show modeled SEDs of starburst galaxies taken from the templates
    of Lagache et al. (2003) magnified by a factor of 75.  The
    magnification factor is taken from the lensing model used by
    Gonzalez et al. (2009) who find a magnification of 25 for the
    first image and of 50 for the second.  The dotted-dashed curve
    corresponds to a starburst galaxy of intrinsic total luminosity
    $L_{\rm tot}=10^{11.8} L_\odot$ redshifted to $z=2.9$.  The solid
    curve shows the SED of a starburst galaxy of $L_{\rm tot} =
    10^{11.9} L_\odot$ redshifted to $z=3.9$.  The data points are the
    24~$\mu$m Spitzer measurement of \cite{GonzalezClowe:2009aa}, the
    250, 350 and 500~$\mu$m BLAST measurements of
    \cite{RexAde:2009aa}, our ~870~$\mu$m LABOCA point, and the AzTEC
    1.1 mm measurement of W08.}
  \label{fig:sed}
\end{figure}

Figure~\ref{fig:sed} shows the spectral energy distribution of
Source~\#1. The measured data points are from from Spitzer, BLAST,
LABOCA~and AzTEC. We compared the models of
\cite{2003MNRAS.338..555L} 
of template starburst galaxies to those data points.  We used a
magnification of 75, assuming that the flux comes from images~A and B
discussed by \cite{GonzalezClowe:2009aa}.  We display the SED of a
starburst galaxy at redshift 2.9, which is the redshift estimate of
\cite{RexAde:2009aa}, with a total luminosity of $10^{11.8} L_\odot$.
The SED of a more redshifted galaxy, at redshift 3.9, with a
luminosity of $10^{11.9} L_\odot$ is in better agreement with the
long-wavelength data. However, it does not fit the 24~$\mu
\mathrm{m}$~value.  Although we have not done a formal fit, it seems
that the current data would require a different template SED.
Alternatively, the uncertainties on the data points may have been
seriously underestimated.

We do not list any Spitzer photometry of source~\#1 because accurate
measurement would require not only processing the raw data but also a
careful subtraction of the foreground elliptical that lies between the
two images of that source. Such a work has been performed by
\cite{GonzalezClowe:2009aa} and we used their quoted values in the
analysis.

\textbf{Source \#2:} This is our second most significant detection at
Gaussian filter value 29.9 or signal-to-noise ratio S/N$\simeq$15. No
previous mm or submm detection of this source has been reported,
although in the map showing the detection of Source \#1 in W08 (their
Fig.~1) there is an indication that they also detected this source.

Fig.~\ref{fig:post} shows that this galaxy has likely counterparts in
all available Spitzer bands. There is a source to the north of the
center of the LABOCA~detection in the 3.6 and 4.5~$\mu \mathrm{m}$~maps that
might be the reason for the apparent elongation of the red contour
towards the north. However, this source is outside the
10\arcsec~search radius and does not show up in the 24~$\mu \mathrm{m}$~map.

\textbf{Source \#3:} A bright, extended Spitzer source is seen to the
southeast of the LABOCA~position. Its large angular size at 8.0~$\mu
\mathrm{m}$~and its distance from the submm position indicates that it
is not the infrared counterpart. It is detected in the 2MASS catalog,
but no redshift is indicated. Two sources are detected within the
10\arcsec~circle. The source closest to the LABOCA~position is
identified in all IRAC band, whereas the other source is only found in
IRAC1 and IRAC2. Both these sources are very faint due to this source
confusion we choose to not list a counterpart. A flux excess is
identified in the MIPS map, but it is not significant enough to be
extracted by SExtractor.

\textbf{Source \#4:} The identification of a counterpart to the submm
source is complicated by the bright foreground source seen in all five
Spitzer bands. This source, at $z=0.097$, is also found in the 2MASS
catalog. In radio observations with the ATCA array,
\cite{Liang:2000hl} detected a source at the position of the
foreground galaxy. The flux density at 1.344 GHz is $0.8\pm 0.05$ mJy
and drops towards higher frequencies. The source is not detected at
4.4 and 8.8 GHz.

Due to the positional offset, it is not very likely that the submm
emission seen in the LABOCA~map is caused by the $z=0.097$ galaxy south
of the submm detection. There is a possibility that a small part of
the total LABOCA~flux density is coming from this foreground source, but
not a substantial fraction. A faint source is detected to the north of
the central position, but its position is slightly outside the
10\arcsec~search radius.

\textbf{Source \#5:}
Two sources with detections in the IRAC 1, 2 and 4 bands are found
close to the central position. A MIPS detection is associated with the
centralmost source, which is situated only 1\arcsec~from the
LABOCA~position. The other source lies 6\arcsec~away which point towards
a counterpart association with the central source. High resolution
radio imaging is needed to completely differentiate between these two
sources, but we list the most central source here as a
counterpart. Another source, to the south-east, is detected only in
IRAC1, and is likely too faint to be a counterpart.

\textbf{Source \#6:} A source is detected in the IRAC 1 and 2 bands
south of the LABOCA~position. The brighter source north-west of the
central position is further out than the search radius.

\textbf{Source \#7:} A significant source is detected in the IRAC 1, 2
and 3 bands west of the center of this source. Its SED is declining
towards the longer wavelength IRAC bands.

\textbf{Source \#8:} A counterpart is identified in all Spitzer bands
to the north of the LABOCA~central position.

\textbf{Source \#9:} Almost 10\arcsec~from the central position lies a
significant source, which is probably too far out and too bright to be
the SMG counterpart. Closer to the central position two IRAC1 sources
are extracted, but their similar flux levels and distance to the
center makes the counterpart association ambiguous. We therefore
choose to not list a counterpart for this SMG.

\textbf{Source \#10:} A bright MIPS source with counterparts in IRAC
1, 2 and 3 is situated close to the LABOCA~central position. This is a
tentative counterpart to the LABOCA~source.

\textbf{Source \#11:} A possible counterpart is detected in all the
Spitzer bands.

\textbf{Source \#15:} The source to the northeast is identified as a
counterpart and exists in all the Spitzer bands.

\textbf{Source \#17:} lies outside the $R=5'$ circle from the center
in a region when the noise is high. Our deboosting algorithm failed in
that case, so we only quote the measured flux density of
$23.6\pm5.9$~mJy. This very bright source is also present in the
AzTEC~1.1~mm map (D. Hughes \& I. Aretxaga, priv. comm.).

\subsection{Number counts}
\label{sec:number-counts}

\begin{figure}[t!]
  \centering
  \includegraphics{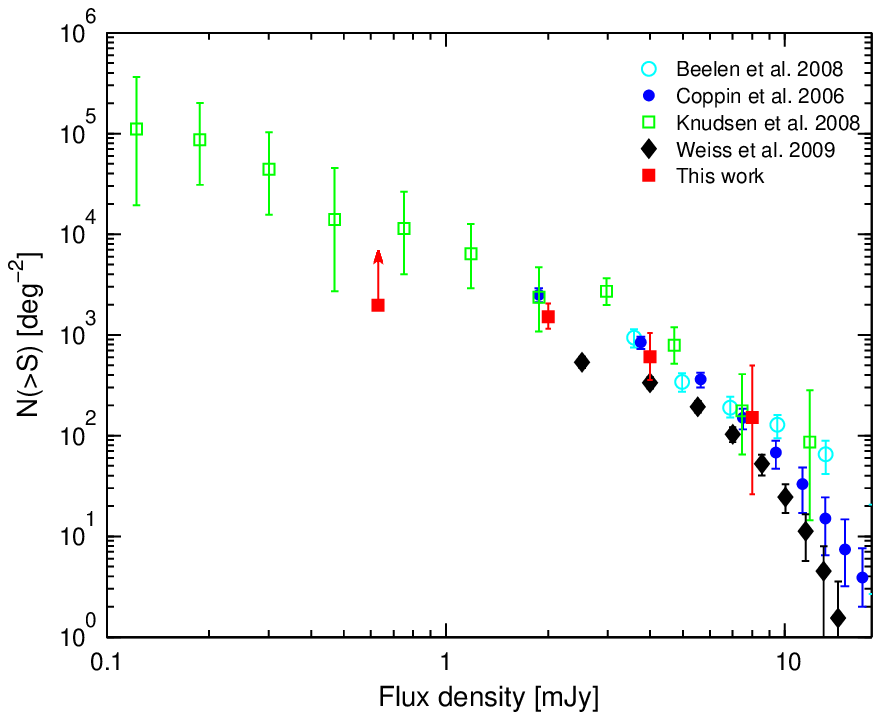}
  \caption{Cumulative number counts derived from the central 10
    arcminutes of the 870~$\mu$m map, after lensing and flux boosting
    correction. As the observations are highly complete, no
    completeness correction has been made (see the discussion in
    Sect.~\ref{sec:completeness} and Fig.~\ref{fig:sim}). For
    comparison, the resulting number counts from the LABOCA~surveys of
    \cite{WeisKovacs:2009ab} and \cite{BeelenOmont:2008aa}, and the
    SCUBA surveys of \cite{CoppinChapin:2006aa} and
    \cite{Knudsenvan-der-Werf:2008aa} are displayed. The arrow on our
    point in the lowest bin indicates a lower limit: we reach the
    sub-mJy regime only because of the one bright source with an
    estimated magnification of 75. The surveyed are for this
    magnification is much smaller than the mean magnification factor
    of 3.3 described in the text, therefore this point only denotes a
    lower limit. }
  \label{fig:nc}
\end{figure}

Figure~\ref{fig:nc} shows the cumulative number counts, defined as the
number density of sources with flux density larger than $S$, denoted
$N(>S)$, derived from our observations and from other studies. There
is good overall agreement between our results and those from previous
LABOCA~and SCUBA surveys, although our surveyed area is not as large as
that of other studies, as reflected by the larger errorbars. 

Because of the gravitational magnification, the surveyed area in the
source plane is smaller than that in the image plane.  We used the
simple lensing model described in Sect.~4.5 to calculate the mean
magnification in the central part of the map used for the analysis. We
calculated a mean magnification factor of 3.3, i.e. the area of the
source plane is 3.3 times smaller than the surveyed area in the image
plane. The magnification of the source fluxes is also taken into
account by using the de-magnified source flux densities listed in
Table~\ref{tab:sources}.

Our counts were inferred from the 13 sources found within the central
10 arcminutes (the region within the red circle shown in
Fig.~\ref{fig:gf}), where the noise is uniform. From our 13 sources we
construct the number counts in four flux-bins. We probe the counts
down to $\sim 0.64$~mJy, a regime which can be investigated only by
utilizing foreground lensing clusters. The number of points per
flux-bin and the counts are displayed in Table~\ref{tab:nc}. We
estimate errorbars for each data point from Poissonian statistics,
using the tabulated values from \cite{Gehrels:1986aa}. 

In Fig.~\ref{fig:nc} we also show the results from the LESS survey
(filled diamonds), which covered an area of $30\arcmin \times
30\arcmin$ of the Extended Chandra Deep Field South to a uniform noise
level of 1.2~mJy/beam \citep{WeisKovacs:2009ab}. That survey has a
noise level comparable to ours, but it covers an area which is 10
times larger. The apparent deficit of galaxies in the LESS has also
been demonstrated in other wavebands (see \cite{WeisKovacs:2009ab} and
references therein). The results of the LABOCA~study of the
protocluster J2142-4423 at $z=2.38$ \citep{BeelenOmont:2008aa} are
plotted in Fig.~\ref{fig:nc} as open circles. The excess seen at
$S>3$~mJy might be an effect of clustering of submm sources in the
protocluster.

We also compare with the SCUBA surveys from \cite{CoppinChapin:2006aa}
(SHADES, filled circles) and \cite{Knudsenvan-der-Werf:2008aa}
(open boxes). The SHADES survey, which covered blank fields,
probes the high-end of the number counts, while the
\citeauthor{Knudsenvan-der-Werf:2008aa} survey targeted 10 lensing
galaxy clusters and reached lower flux levels. Although our surveyed
area is similar to that of \cite{Knudsenvan-der-Werf:2008aa}, we find
only one sub-mJy source compared to the seven found by
\citeauthor{Knudsenvan-der-Werf:2008aa} This is because their survey
covers a larger area of high magnification. 

\begin{table}[t]
  \centering
  \caption{Cumulative number counts based on the central 10~arcminutes of our map.}
  \begin{tabular}[h]{c c c c}
    \hline
    \hline
    $S_{870}$ & $N(>S)$ & $N_{\mathrm{sources}}$\tablefootmark{a} &
    $N_{\mathrm{per\,bin}}$ \tablefootmark{b} \\
     mJy & deg$^{-2}$ \\
    \hline
    \vspace{1ex}
    0.64 & $>1970$ & 13 & 3 \\
    \vspace{1ex}
    2.0 & $1510_{-360}^{+540}$ & 10 & 6 \\
    \vspace{1ex}
    4.0 & $610_{-250}^{+440}$ & 4 & 3 \\
    \vspace{1ex}
    8.0 & $150_{-130}^{+350}$ & 1 & 1 \\
    \hline   
  \end{tabular}
  \tablefoot{\tablefoottext{a} Cumulative number of
    sources. \tablefoottext{b} The number of sources contributing to
    a bin (used to estimate the uncertainty).}
  \label{tab:nc}
\end{table}

\subsection{Resolving the cosmic infrared background}
\label{sec:resolv-cosm-infr}

It is interesting to investigate how much of the cosmic infrared
background (CIB) flux is resolved in our LABOCA~observation. The CIB was
discovered by the COBE instruments FIRAS and DIRBE
\citep{FixsenDwek:1998aa,HauserArendt:1998aa}. Those experiments
detected an isotropic infrared background signal that was thought to
originate from the integrated effect of star formation in the history
of the universe \citep{DwekArendt:1998aa}. But the coarse resolution
of the COBE satellite made it impossible to resolve the sources
responsible for the background signal. The detection of those sources
had to wait for the invention of mm and submm bolometer cameras on
large ground-based telescopes, resulting in angular resolutions of
tens of arcseconds.

Because of the steep number counts, most of the flux coming from submm
galaxies originates in low-flux sources.  Therefore, observations of
lensing foreground clusters that probe the faintest number counts
potentially will resolve a large fraction of the CIB.

We use the central region of 10\arcmin~diameter for this
calculation. The amount of CIB flux in that area, at 870~$\mu \mathrm{m}$~is
estimated from Eq.~1 of \cite{DwekArendt:1998aa} and yields $\sim 900$
mJy. The total flux in the sources extracted from the LABOCA~map within
that region (the same sources that are used for the number counts, but
without the flux correction due to magnification) is $124\pm 4.0$
mJy. This calculation is valid because gravitational lensing preserves
surface brightness. Thus, 14\% of the CIB has been resolved in the
LABOCA observations.

\subsection{Residual SZ emission}

Although the data have been filtered to remove the extended emission,
some level of residual SZ emission may be present.  Close inspection
of the central region of the map shows extended emission between the
two brighter sources, Source~\#1 and \#2.  We note that W08 (their
Fig.~1) see a similar extended structure around the same central
sources.  In order to set an upper limit to the amount of residual SZ
signal present in the region of the map covered by the Bullet cluster
(which corresponds roughly to the area of a disk with an outer radius
of about 2 arcminutes), we calculated the difference between the
excess flux density in that region (estimated from
Fig.~\ref{fig:hist}) and the flux density in resolved point sources in
the same region. Removing the positive noise contribution estimated
from the jackknife noise maps we find that 40~mJy residual SZ emission
might be present in the final map.

We can use the results of the APEX-SZ observations of the decrement at
150~GHz by \cite{2009ApJ...701...42H} 
to estimate the SZ flux density across the sky area covered by the
Bullet Cluster. \cite{2009ApJ...701...42H} fitted an elliptical beta
model using an X-ray prior on $\beta = 1.04^{+0.16}_{-0.10}$ and found
a core radius $\theta_c=142\pm18''$, an axial ratio of $0.89\pm0.07$
and a central Compton parameter $y_0=3.31 \pm 0.30 \times 10^{-4}$.
Integrating the beta model over the central region with a maximum
radius of 2 arcminutes, and taking for simplicity a spherical model
with the corresponding core radius, we estimate that the total SZ flux
density from that area is of the order of 400~mJy. Therefore, we
cannot exclude that a maximum of 10\% of the total SZ flux of the
Bullet Cluster may remain in the map in the form of extended emission.

\section{Conclusions}

Continuum observations at 870~$\mu \mathrm{m}$~of the Bullet Cluster have been
presented. The data were filtered to remove large-scale signal such as
the Sunyaev--Zeldovich~increment from the cluster. The main results are summarized
below:

\begin{itemize}
\item Seventeen submm sources with signal-to-noise ratios larger than
  4 were detected in the map. Their measured fluxes densities range
  from 4.6 to 48~mJy. For each source but one, which lies in a noisy
  area, we calculated the value of the flux density corrected for the
  flux boosting due to confusion noise and for the gravitational
  magnification by the cluster.
\item The brightest submm source coincides with a previously reported
  galaxy at an estimated redshift of 2.7--2.9 detected by Spitzer and
  the AzTEC~and BLAST experiments. With its flux density of 48~mJy it
  is one of the brightest submm galaxy ever
  detected. After correction for gravitational magnification $|\mu|$,
  the intrinsic flux of the source is about $0.64 \,
  (|\mu|/75)^{-1}$~mJy.

\item We found reliable infrared counterparts for nine of the submm
  sources. An infrared color-color analysis suggests that they have
  starburst-dominated spectral energy distributions.

 \item The cumulative number counts derived from the observations
   agree well with those from other surveys.

 \item The observations resolve 14\% of the cosmic infrared background
   radiation at 870~$\mu \mathrm{m}$, in a sky area of $\sim 78.5\,
   \mathrm{arcmin}^2$.

\end{itemize}

We plan to apply a similar analysis to several other cluster fields
for which we have LABOCA~data. The results presented here will also be
used to remove the contribution of the submm sources and obtain a map
of the SZ increment at 870~$\mu \mathrm{m}$~in the Bullet Cluster~with
sub-arcminute angular resolution. Although the SZ decrement of the
Bullet Cluster has been observed using several instruments (e.g., SEST
by \citealt{Andreani:1999bl}, APEX-SZ by
\citealt{2009ApJ...701...42H}, ACT by
\citealt{HincksAcquaviva:2009aa}, and recently the South Pole
Telescope by \citealt{PlaggeBenson:2009aa}) the only observation of
the increment was done using ACBAR at 275~GHz, with an angular
resolution of 4.5\arcmin~\citep{Gomez:2004dz}. Removal of lensed
background sources is not a trivial task, since it can also bias the
SZ measurement \citep{LoebRefregier:1997aa}. In the case of the Bullet
Cluster, the extremely bright lensed submm source close to the center
of the cluster field has an integrated flux density comparable to that
of the SZ flux from the central region of the cluster. A careful
analysis must be performed to recover the SZ increment in such a
system, and may require a joint analysis using measurements in other
wavebands.

\begin{acknowledgements}

  We thank the APEX staff for excellent support during the
  observations. CH and DJ thank the Swedish Research Council for
  financial support. MS, KB, and FB acknowledge support from a grant
  within the DFG Priority Program 1177. HQ acknowledges partial
  support from the FONDAP Centro de Astrof{\'i}sica. We thank
  Alexandre Beelen, John H. Black, David Hughes and Itziar Aretxaga
  for stimulating discussions, and Florian Pacaud for providing us
  with the XMM map of the Bullet Cluster shown as contours in
  Fig.~7. We are grateful to Attila Kov\'acs for writing {\tt
    Minicrush}~ and being helpful to us during the data reduction. We
  thank the referee for a careful reading of the manuscript and useful
  comments.

  This work is based in part on observations made with the Spitzer
  Space Telescope, which is operated by the Jet Propulsion Laboratory,
  California Institute of Technology under a contract with NASA.
\end{acknowledgements}

\bibliographystyle{aa}
\bibliography{13833}

\begin{thebibliography}{57}
\expandafter\ifx\csname natexlab\endcsname\relax\def\natexlab#1{#1}\fi

\bibitem[{{Andreani} {et~al.}(1999){Andreani}, {B{\"o}hringer}, {dall'Oglio},
  {Martinis}, {Shaver}, {Lemke}, {Nyman}, {Booth}, {Pizzo}, {Whyborn},
  {Tanaka}, \& {Liang}}]{Andreani:1999bl}
{Andreani}, P., {B{\"o}hringer}, H., {dall'Oglio}, G., {et~al.} 1999, \apj,
  513, 23

\bibitem[{{Austermann} {et~al.}(2010){Austermann}, {Dunlop}, {Perera}, {Scott},
  {Wilson}, {Aretxaga}, {Hughes}, {Almaini}, {Chapin}, {Chapman}, {Cirasuolo},
  {Clements}, {Coppin}, {Dunne}, {Dye}, {Eales}, {Egami}, {Farrah}, {Ferrusca},
  {Flynn}, {Haig}, {Halpern}, {Ibar}, {Ivison}, {van Kampen}, {Kang}, {Kim},
  {Lacey}, {Lowenthal}, {Mauskopf}, {McLure}, {Mortier}, {Negrello}, {Oliver},
  {Peacock}, {Pope}, {Rawlings}, {Rieke}, {Roseboom}, {Rowan-Robinson},
  {Scott}, {Serjeant}, {Smail}, {Swinbank}, {Stevens}, {Velazquez}, {Wagg}, \&
  {Yun}}]{AustermannDunlop:2010aa}
{Austermann}, J.~E., {Dunlop}, J.~S., {Perera}, T.~A., {et~al.} 2010, \mnras,
  401, 160

\bibitem[{{Beelen} {et~al.}(2008){Beelen}, {Omont}, {Bavouzet}, {Kov{\'a}cs},
  {Lagache}, {De Breuck}, {Weiss}, {Menten}, {Colbert}, {Dole}, {Siringo}, \&
  {Kreysa}}]{BeelenOmont:2008aa}
{Beelen}, A., {Omont}, A., {Bavouzet}, N., {et~al.} 2008, \aap, 485, 645

\bibitem[{{Bertin} \& {Arnouts}(1996)}]{1996A&AS..117..393B}
{Bertin}, E. \& {Arnouts}, S. 1996, \aaps, 117, 393

\bibitem[{{Bertoldi} {et~al.}(2007){Bertoldi}, {Carilli}, {Aravena},
  {Schinnerer}, {Voss}, {Smolcic}, {Jahnke}, {Scoville}, {Blain}, {Menten},
  {Lutz}, {Brusa}, {Taniguchi}, {Capak}, {Mobasher}, {Lilly}, {Thompson},
  {Aussel}, {Kreysa}, {Hasinger}, {Aguirre}, {Schlaerth}, \&
  {Koekemoer}}]{BertoldiCarilli:2007aa}
{Bertoldi}, F., {Carilli}, C., {Aravena}, M., {et~al.} 2007, \apjs, 172, 132

\bibitem[{{Blain}(1997)}]{1997MNRAS.290..553B}
{Blain}, A.~W. 1997, \mnras, 290, 553

\bibitem[{{Blain}(1998)}]{Blain:1998aa}
{Blain}, A.~W. 1998, \mnras, 297, 502

\bibitem[{{Blain} \& {Longair}(1993)}]{1993MNRAS.264..509B}
{Blain}, A.~W. \& {Longair}, M.~S. 1993, \mnras, 264, 509

\bibitem[{{Blain} {et~al.}(2002){Blain}, {Smail}, {Ivison}, {Kneib}, \&
  {Frayer}}]{BlainSmail:2002aa}
{Blain}, A.~W., {Smail}, I., {Ivison}, R.~J., {Kneib}, J.-P., \& {Frayer},
  D.~T. 2002, \physrep, 369, 111

\bibitem[{{Borys} {et~al.}(2003){Borys}, {Chapman}, {Halpern}, \&
  {Scott}}]{BorysChapman:2003aa}
{Borys}, C., {Chapman}, S., {Halpern}, M., \& {Scott}, D. 2003, \mnras, 344,
  385

\bibitem[{{Brada{\v c}} {et~al.}(2006){Brada{\v c}}, {Clowe}, {Gonzalez},
  {Marshall}, {Forman}, {Jones}, {Markevitch}, {Randall}, {Schrabback}, \&
  {Zaritsky}}]{Bradac:2006fj}
{Brada{\v c}}, M., {Clowe}, D., {Gonzalez}, A.~H., {et~al.} 2006, \apj, 652,
  937

\bibitem[{{Chapman} {et~al.}(2002){Chapman}, {Scott}, {Borys}, \&
  {Fahlman}}]{2002MNRAS.330...92C}
{Chapman}, S.~C., {Scott}, D., {Borys}, C., \& {Fahlman}, G.~G. 2002, \mnras,
  330, 92

\bibitem[{{Clowe} {et~al.}(2006){Clowe}, {Brada{\v c}}, {Gonzalez},
  {Markevitch}, {Randall}, {Jones}, \& {Zaritsky}}]{Clowe:2006th}
{Clowe}, D., {Brada{\v c}}, M., {Gonzalez}, A.~H., {et~al.} 2006, \apjl, 648,
  L109

\bibitem[{{Condon}(1974)}]{Condon:1974aa}
{Condon}, J.~J. 1974, \apj, 188, 279

\bibitem[{{Coppin} {et~al.}(2006){Coppin}, {Chapin}, {Mortier}, {Scott},
  {Borys}, {Dunlop}, {Halpern}, {Hughes}, {Pope}, {Scott}, {Serjeant}, {Wagg},
  {Alexander}, {Almaini}, {Aretxaga}, {Babbedge}, {Best}, {Blain}, {Chapman},
  {Clements}, {Crawford}, {Dunne}, {Eales}, {Edge}, {Farrah}, {Gazta{\~n}aga},
  {Gear}, {Granato}, {Greve}, {Fox}, {Ivison}, {Jarvis}, {Jenness}, {Lacey},
  {Lepage}, {Mann}, {Marsden}, {Martinez-Sansigre}, {Oliver}, {Page},
  {Peacock}, {Pearson}, {Percival}, {Priddey}, {Rawlings}, {Rowan-Robinson},
  {Savage}, {Seigar}, {Sekiguchi}, {Silva}, {Simpson}, {Smail}, {Stevens},
  {Takagi}, {Vaccari}, {van Kampen}, \& {Willott}}]{CoppinChapin:2006aa}
{Coppin}, K., {Chapin}, E.~L., {Mortier}, A.~M.~J., {et~al.} 2006, \mnras, 372,
  1621

\bibitem[{{Coppin} {et~al.}(2005){Coppin}, {Halpern}, {Scott}, {Borys}, \&
  {Chapman}}]{CoppinHalpern:2005aa}
{Coppin}, K., {Halpern}, M., {Scott}, D., {Borys}, C., \& {Chapman}, S. 2005,
  \mnras, 357, 1022

\bibitem[{{Cowie} {et~al.}(2002){Cowie}, {Barger}, \&
  {Kneib}}]{CowieBarger:2002aa}
{Cowie}, L.~L., {Barger}, A.~J., \& {Kneib}, J.-P. 2002, \aj, 123, 2197

\bibitem[{{Dwek} {et~al.}(1998){Dwek}, {Arendt}, {Hauser}, {Fixsen}, {Kelsall},
  {Leisawitz}, {Pei}, {Wright}, {Mather}, {Moseley}, {Odegard}, {Shafer},
  {Silverberg}, \& {Weiland}}]{DwekArendt:1998aa}
{Dwek}, E., {Arendt}, R.~G., {Hauser}, M.~G., {et~al.} 1998, \apj, 508, 106

\bibitem[{{Fazio} {et~al.}(2004){Fazio}, {Hora}, {Allen}, {Ashby}, {Barmby},
  {Deutsch}, {Huang}, {Kleiner}, {Marengo}, {Megeath}, {Melnick}, {Pahre},
  {Patten}, {Polizotti}, {Smith}, {Taylor}, {Wang}, {Willner}, {Hoffmann},
  {Pipher}, {Forrest}, {McMurty}, {McCreight}, {McKelvey}, {McMurray}, {Koch},
  {Moseley}, {Arendt}, {Mentzell}, {Marx}, {Losch}, {Mayman}, {Eichhorn},
  {Krebs}, {Jhabvala}, {Gezari}, {Fixsen}, {Flores}, {Shakoorzadeh}, {Jungo},
  {Hakun}, {Workman}, {Karpati}, {Kichak}, {Whitley}, {Mann}, {Tollestrup},
  {Eisenhardt}, {Stern}, {Gorjian}, {Bhattacharya}, {Carey}, {Nelson},
  {Glaccum}, {Lacy}, {Lowrance}, {Laine}, {Reach}, {Stauffer}, {Surace},
  {Wilson}, {Wright}, {Hoffman}, {Domingo}, \& {Cohen}}]{FazioHora:2004aa}
{Fazio}, G.~G., {Hora}, J.~L., {Allen}, L.~E., {et~al.} 2004, \apjs, 154, 10

\bibitem[{{Fixsen} {et~al.}(1998){Fixsen}, {Dwek}, {Mather}, {Bennett}, \&
  {Shafer}}]{FixsenDwek:1998aa}
{Fixsen}, D.~J., {Dwek}, E., {Mather}, J.~C., {Bennett}, C.~L., \& {Shafer},
  R.~A. 1998, \apj, 508, 123

\bibitem[{{Gehrels}(1986)}]{Gehrels:1986aa}
{Gehrels}, N. 1986, \apj, 303, 336

\bibitem[{{Gomez} {et~al.}(2004){Gomez}, {Romer}, {Peterson}, {Chase},
  {Runyan}, {Holzapfel}, {Kuo}, {Newcomb}, {Ruhl}, {Goldstein}, \&
  {Lange}}]{Gomez:2004dz}
{Gomez}, P., {Romer}, A.~K., {Peterson}, J.~B., {et~al.} 2004, in AIP Conf.
  Proc. 703: Plasmas in the Laboratory and in the Universe: New Insights and
  New Challenges, ed. G.~{Bertin}, D.~{Farina}, \& R.~{Pozzoli}, 361--366

\bibitem[{{Gonzalez} {et~al.}(2009){Gonzalez}, {Clowe}, {Brada{\v c}},
  {Zaritsky}, {Jones}, \& {Markevitch}}]{GonzalezClowe:2009aa}
{Gonzalez}, A.~H., {Clowe}, D., {Brada{\v c}}, M., {et~al.} 2009, \apj, 691,
  525

\bibitem[{{G{\"u}sten} {et~al.}(2006){G{\"u}sten}, {Nyman}, {Schilke},
  {Menten}, {Cesarsky}, \& {Booth}}]{Gusten:2006ak}
{G{\"u}sten}, R., {Nyman}, L.~{\AA}., {Schilke}, P., {et~al.} 2006, \aap, 454,
  L13

\bibitem[{{Hainline} {et~al.}(2009){Hainline}, {Blain}, {Smail}, {Frayer},
  {Chapman}, {Ivison}, \& {Alexander}}]{HainlineBlain:2009aa}
{Hainline}, L.~J., {Blain}, A.~W., {Smail}, I., {et~al.} 2009, \apj, 699, 1610

\bibitem[{{Halverson} {et~al.}(2009){Halverson}, {Lanting}, {Ade}, {Basu},
  {Bender}, {Benson}, {Bertoldi}, {Cho}, {Chon}, {Clarke}, {Dobbs}, {Ferrusca},
  {G{\"u}sten}, {Holzapfel}, {Kov{\'a}cs}, {Kennedy}, {Kermish}, {Kneissl},
  {Lee}, {Lueker}, {Mehl}, {Menten}, {Muders}, {Nord}, {Pacaud}, {Plagge},
  {Reichardt}, {Richards}, {Schaaf}, {Schilke}, {Schuller}, {Schwan},
  {Spieler}, {Tucker}, {Weiss}, \& {Zahn}}]{2009ApJ...701...42H}
{Halverson}, N.~W., {Lanting}, T., {Ade}, P.~A.~R., {et~al.} 2009, \apj, 701,
  42

\bibitem[{{Hauser} {et~al.}(1998){Hauser}, {Arendt}, {Kelsall}, {Dwek},
  {Odegard}, {Weiland}, {Freudenreich}, {Silverberg}, {Moseley}, {Pei},
  {Lubin}, {Mather}, {Shafer}, {Smoot}, {Weiss}, {Wilkinson}, \&
  {Wright}}]{HauserArendt:1998aa}
{Hauser}, M.~G., {Arendt}, R.~G., {Kelsall}, T., {et~al.} 1998, \apj, 508, 25

\bibitem[{{Hincks} {et~al.}(2009){Hincks}, {Acquaviva}, {Ade}, {Aguirre},
  {Amiri}, {Appel}, {Barrientos}, {Battistelli}, {Bond}, {Brown}, {Burger},
  {Chervenak}, {Das}, {Devlin}, {Dicker}, {Doriese}, {Dunkley}, {D{\"u}nner},
  {Essinger-Hileman}, {Fisher}, {Fowler}, {Hajian}, {Halpern}, {Hasselfield},
  {Hern{\'a}ndez-Monteagudo}, {Hilton}, {Hilton}, {Hlozek}, {Huffenberger},
  {Hughes}, {Hughes}, {Infante}, {Irwin}, {Jimenez}, {Juin}, {Kaul}, {Klein},
  {Kosowsky}, {Lau}, {Limon}, {Lin}, {Lupton}, {Marriage}, {Marsden},
  {Martocci}, {Mauskopf}, {Menanteau}, {Moodley}, {Moseley}, {Netterfield},
  {Niemack}, {Nolta}, {Page}, {Parker}, {Partridge}, {Quintana}, {Reid},
  {Sehgal}, {Sievers}, {Spergel}, {Staggs}, {Stryzak}, {Swetz}, {Switzer},
  {Thornton}, {Trac}, {Tucker}, {Verde}, {Warne}, {Wilson}, {Wollack}, \&
  {Zhao}}]{HincksAcquaviva:2009aa}
{Hincks}, A.~D., {Acquaviva}, V., {Ade}, P., {et~al.} 2009, ArXiv e-prints

\bibitem[{{Hogg} \& {Turner}(1998)}]{HoggTurner:1998aa}
{Hogg}, D.~W. \& {Turner}, E.~L. 1998, \pasp, 110, 727

\bibitem[{{Ivison} {et~al.}(2007){Ivison}, {Greve}, {Dunlop}, {Peacock},
  {Egami}, {Smail}, {Ibar}, {van Kampen}, {Aretxaga}, {Babbedge}, {Biggs},
  {Blain}, {Chapman}, {Clements}, {Coppin}, {Farrah}, {Halpern}, {Hughes},
  {Jarvis}, {Jenness}, {Jones}, {Mortier}, {Oliver}, {Papovich},
  {P{\'e}rez-Gonz{\'a}lez}, {Pope}, {Rawlings}, {Rieke}, {Rowan-Robinson},
  {Savage}, {Scott}, {Seigar}, {Serjeant}, {Simpson}, {Stevens}, {Vaccari},
  {Wagg}, \& {Willott}}]{IvisonGreve:2007aa}
{Ivison}, R.~J., {Greve}, T.~R., {Dunlop}, J.~S., {et~al.} 2007, \mnras, 380,
  199

\bibitem[{{Ivison} {et~al.}(2004){Ivison}, {Greve}, {Serjeant}, {Bertoldi},
  {Egami}, {Mortier}, {Alonso-Herrero}, {Barmby}, {Bei}, {Dole}, {Engelbracht},
  {Fazio}, {Frayer}, {Gordon}, {Hines}, {Huang}, {Le Floc'h}, {Misselt},
  {Miyazaki}, {Morrison}, {Papovich}, {P{\'e}rez-Gonz{\'a}lez}, {Rieke},
  {Rieke}, {Rigby}, {Rigopoulou}, {Smail}, {Wilson}, \&
  {Willner}}]{IvisonGreve:2004aa}
{Ivison}, R.~J., {Greve}, T.~R., {Serjeant}, S., {et~al.} 2004, \apjs, 154, 124

\bibitem[{{Knudsen} {et~al.}(2008){Knudsen}, {van der Werf}, \&
  {Kneib}}]{Knudsenvan-der-Werf:2008aa}
{Knudsen}, K.~K., {van der Werf}, P.~P., \& {Kneib}, J.-P. 2008, \mnras, 384,
  1611

\bibitem[{{Kov{\'a}cs}(2008)}]{2008SPIE.7020E..45K}
{Kov{\'a}cs}, A. 2008, in Presented at the Society of Photo-Optical
  Instrumentation Engineers (SPIE) Conference, Vol. 7020, Society of
  Photo-Optical Instrumentation Engineers (SPIE) Conference Series

\bibitem[{{Lagache} {et~al.}(2003){Lagache}, {Dole}, \&
  {Puget}}]{2003MNRAS.338..555L}
{Lagache}, G., {Dole}, H., \& {Puget}, J. 2003, \mnras, 338, 555

\bibitem[{{Liang} {et~al.}(2000){Liang}, {Hunstead}, {Birkinshaw}, \&
  {Andreani}}]{Liang:2000hl}
{Liang}, H., {Hunstead}, R.~W., {Birkinshaw}, M., \& {Andreani}, P. 2000, \apj,
  544, 686

\bibitem[{{Loeb} \& {Refregier}(1997)}]{LoebRefregier:1997aa}
{Loeb}, A. \& {Refregier}, A. 1997, \apjl, 476, L59+

\bibitem[{{Markevitch} {et~al.}(2002){Markevitch}, {Gonzalez}, {David},
  {Vikhlinin}, {Murray}, {Forman}, {Jones}, \& {Tucker}}]{Markevitch:2002lr}
{Markevitch}, M., {Gonzalez}, A.~H., {David}, L., {et~al.} 2002, \apjl, 567,
  L27

\bibitem[{{Nord} {et~al.}(2009){Nord}, {Basu}, {Pacaud}, {Ade}, {Bender},
  {Benson}, {Bertoldi}, {Cho}, {Chon}, {Clarke}, {Dobbs}, {Ferrusca},
  {Halverson}, {Holzapfel}, {Horellou}, {Johansson}, {Kennedy}, {Kermish},
  {Kneissl}, {Lanting}, {Lee}, {Lueker}, {Mehl}, {Menten}, {Plagge},
  {Reichardt}, {Richards}, {Schaaf}, {Schwan}, {Spieler}, {Tucker}, {Weiss}, \&
  {Zahn}}]{2009A&A...506..623N}
{Nord}, M., {Basu}, K., {Pacaud}, F., {et~al.} 2009, \aap, 506, 623

\bibitem[{{Plagge} {et~al.}(2009){Plagge}, {Benson}, {Ade}, {Aird}, {Bleem},
  {Carlstrom}, {Chang}, {Cho}, {Crawford}, {Crites}, {de Haan}, {Dobbs},
  {George}, {Hall}, {Halverson}, {Holder}, {Holzapfel}, {Hrubes}, {Joy},
  {Keisler}, {Knox}, {Lee}, {Leitch}, {Lueker}, {Marrone}, {McMahon}, {Mehl},
  {Meyer}, {Mohr}, {Montroy}, {Padin}, {Pryke}, {Reichardt}, {Ruhl},
  {Schaffer}, {Shaw}, {Shirokoff}, {Spieler}, {Stalder}, {Staniszewski},
  {Stark}, {Vanderlinde}, {Vieira}, {Williamson}, \&
  {Zahn}}]{PlaggeBenson:2009aa}
{Plagge}, T., {Benson}, B.~A., {Ade}, P.~A.~R., {et~al.} 2009, ArXiv e-prints

\bibitem[{{Pope} {et~al.}(2006){Pope}, {Scott}, {Dickinson}, {Chary},
  {Morrison}, {Borys}, {Sajina}, {Alexander}, {Daddi}, {Frayer}, {MacDonald},
  \& {Stern}}]{PopeScott:2006aa}
{Pope}, A., {Scott}, D., {Dickinson}, M., {et~al.} 2006, \mnras, 370, 1185

\bibitem[{{Rex} {et~al.}(2009){Rex}, {Ade}, {Aretxaga}, {Bock}, {Chapin},
  {Devlin}, {Dicker}, {Griffin}, {Gundersen}, {Halpern}, {Hargrave}, {Hughes},
  {Klein}, {Marsden}, {Martin}, {Mauskopf}, {Monta{\~n}a}, {Netterfield},
  {Olmi}, {Pascale}, {Patanchon}, {Scott}, {Semisch}, {Thomas}, {Truch},
  {Tucker}, {Tucker}, {Viero}, \& {Wiebe}}]{RexAde:2009aa}
{Rex}, M., {Ade}, P.~A.~R., {Aretxaga}, I., {et~al.} 2009, \apj, 703, 348

\bibitem[{{Rieke} {et~al.}(2004){Rieke}, {Young}, {Engelbracht}, {Kelly},
  {Low}, {Haller}, {Beeman}, {Gordon}, {Stansberry}, {Misselt}, {Cadien},
  {Morrison}, {Rivlis}, {Latter}, {Noriega-Crespo}, {Padgett}, {Stapelfeldt},
  {Hines}, {Egami}, {Muzerolle}, {Alonso-Herrero}, {Blaylock}, {Dole}, {Hinz},
  {Le Floc'h}, {Papovich}, {P{\'e}rez-Gonz{\'a}lez}, {Smith}, {Su}, {Bennett},
  {Frayer}, {Henderson}, {Lu}, {Masci}, {Pesenson}, {Rebull}, {Rho}, {Keene},
  {Stolovy}, {Wachter}, {Wheaton}, {Werner}, \& {Richards}}]{RiekeYoung:2004aa}
{Rieke}, G.~H., {Young}, E.~T., {Engelbracht}, C.~W., {et~al.} 2004, \apjs,
  154, 25

\bibitem[{{Sanders} \& {Mirabel}(1996)}]{SandersMirabel:1996aa}
{Sanders}, D.~B. \& {Mirabel}, I.~F. 1996, \araa, 34, 749

\bibitem[{{Schechter}(1976)}]{Schechter:1976aa}
{Schechter}, P. 1976, \apj, 203, 297

\bibitem[{{Schneider} {et~al.}(1992){Schneider}, {Ehlers}, \&
  {Falco}}]{1992grle.book.....S}
{Schneider}, P., {Ehlers}, J., \& {Falco}, E.~E. 1992, {Gravitational Lenses},
  ed. E.~J. . F. E.~E. Schneider, P.

\bibitem[{{Schneider} \& {Weiss}(1986)}]{1986A&A...164..237S}
{Schneider}, P. \& {Weiss}, A. 1986, \aap, 164, 237

\bibitem[{{Scott} {et~al.}(2008){Scott}, {Austermann}, {Perera}, {Wilson},
  {Aretxaga}, {Bock}, {Hughes}, {Kang}, {Kim}, {Mauskopf}, {Sanders},
  {Scoville}, \& {Yun}}]{ScottAustermann:2008aa}
{Scott}, K.~S., {Austermann}, J.~E., {Perera}, T.~A., {et~al.} 2008, \mnras,
  385, 2225

\bibitem[{{Scott} {et~al.}(2006){Scott}, {Dunlop}, \&
  {Serjeant}}]{2006MNRAS.370.1057S}
{Scott}, S.~E., {Dunlop}, J.~S., \& {Serjeant}, S. 2006, \mnras, 370, 1057

\bibitem[{{Serjeant} {et~al.}(2003){Serjeant}, {Dunlop}, {Mann},
  {Rowan-Robinson}, {Hughes}, {Efstathiou}, {Blain}, {Fox}, {Ivison},
  {Jenness}, {Lawrence}, {Longair}, {Oliver}, \&
  {Peacock}}]{SerjeantDunlop:2003aa}
{Serjeant}, S., {Dunlop}, J.~S., {Mann}, R.~G., {et~al.} 2003, \mnras, 344, 887

\bibitem[{{Shin} \& {Evans}(2008)}]{2008MNRAS.390..505S}
{Shin}, E.~M. \& {Evans}, N.~W. 2008, \mnras, 390, 505

\bibitem[{{Siringo} {et~al.}(2009){Siringo}, {Kreysa}, {Kov{\'a}cs},
  {Schuller}, {Wei{\ss}}, {Esch}, {Gem{\"u}nd}, {Jethava}, {Lundershausen},
  {Colin}, {G{\"u}sten}, {Menten}, {Beelen}, {Bertoldi}, {Beeman}, \&
  {Haller}}]{SiringoKreysa:2009aa}
{Siringo}, G., {Kreysa}, E., {Kov{\'a}cs}, A., {et~al.} 2009, \aap, 497, 945

\bibitem[{{Smail} {et~al.}(1997){Smail}, {Ivison}, \&
  {Blain}}]{SmailIvison:1997aa}
{Smail}, I., {Ivison}, R.~J., \& {Blain}, A.~W. 1997, \apjl, 490, L5+

\bibitem[{{Smail} {et~al.}(1998){Smail}, {Ivison}, {Blain}, \&
  {Kneib}}]{1998ApJ...507L..21S}
{Smail}, I., {Ivison}, R.~J., {Blain}, A.~W., \& {Kneib}, J.-P. 1998, \apjl,
  507, L21

\bibitem[{{Smail} {et~al.}(2000){Smail}, {Ivison}, {Owen}, {Blain}, \&
  {Kneib}}]{SmailIvison:2000aa}
{Smail}, I., {Ivison}, R.~J., {Owen}, F.~N., {Blain}, A.~W., \& {Kneib}, J.-P.
  2000, \apj, 528, 612

\bibitem[{{Springel} \& {Farrar}(2007)}]{SpringelFarrar:2007aa}
{Springel}, V. \& {Farrar}, G.~R. 2007, \mnras, 380, 911

\bibitem[{{Wei{\ss}} {et~al.}(2009){Wei{\ss}}, {Kov{\'a}cs}, {Coppin}, {Greve},
  {Walter}, {Smail}, {Dunlop}, {Knudsen}, {Alexander}, {Bertoldi}, {Brandt},
  {Chapman}, {Cox}, {Dannerbauer}, {De Breuck}, {Gawiser}, {Ivison}, {Lutz},
  {Menten}, {Koekemoer}, {Kreysa}, {Kurczynski}, {Rix}, {Schinnerer}, \& {van
  der Werf}}]{WeisKovacs:2009ab}
{Wei{\ss}}, A., {Kov{\'a}cs}, A., {Coppin}, K., {et~al.} 2009, \apj, 707, 1201

\bibitem[{{Wilson} {et~al.}(2008){Wilson}, {Hughes}, {Aretxaga}, {Ezawa},
  {Austermann}, {Doyle}, {Ferrusca}, {Hern{\'a}ndez-Curiel}, {Kawabe},
  {Kitayama}, {Kohno}, {Kuboi}, {Matsuo}, {Mauskopf}, {Murakoshi},
  {Monta{\~n}a}, {Natarajan}, {Oshima}, {Ota}, {Perera}, {Rand}, {Scott},
  {Tanaka}, {Tsuboi}, {Williams}, {Yamaguchi}, \& {Yun}}]{2008MNRAS.390.1061W}
{Wilson}, G.~W., {Hughes}, D.~H., {Aretxaga}, I., {et~al.} 2008, \mnras, 390,
  1061

\end{thebibliography}

\appendix

\section{Flux deboosting}
\label{sec:flux-deboosting}

\begin{figure*}[h!]
  \centering
  \includegraphics[width=17.7cm]{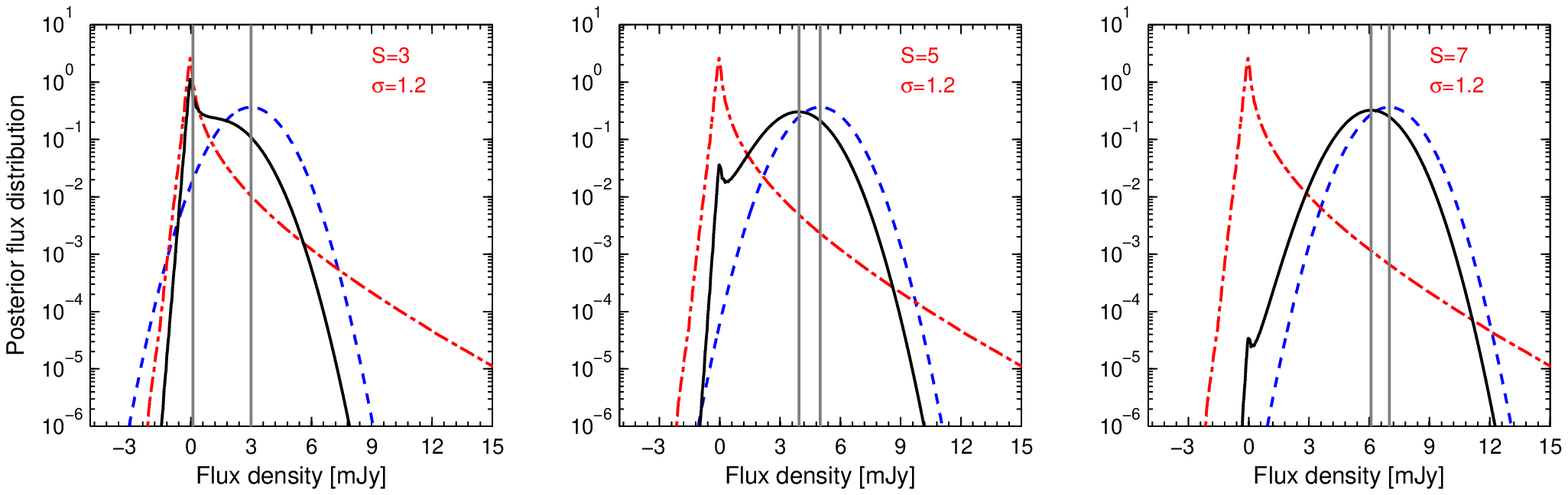}
  \caption{\emph{Solid curves}: Posterior flux distributions for
    measured flux densities of 3, 5 and 7 mJy and a noise level of 1.2
    mJy. \emph{Dashed curves}: Measured flux distributions,
    modelled as Gaussian functions. \emph{Dotted-dashed curve}:
    Prior flux density distributions estimated from noise-free
    simulations, populating maps with a Schechter flux distribution
    (Eq.~(\ref{eq:8})). The prior (red dotted-dashed) is identical in
    all three panels. The posterior flux distribution is the result of
    multiplying the blue dashed curve by the red dot-dashed curve and
    normalizing. The two vertical lines in each panel show the measured
    flux (rightmost line) and the maximum of the posterior flux
    distribution (leftmost line). The difference between the two lines
    indicates the amount of flux boosting.}
  \label{fig:pfd}
\end{figure*}

This appendix describes the flux deboosting algorithm used in this
work, following the prescription in
\citet[Sec. 5.1]{ScottAustermann:2008aa}, whose notation we use in the
rest of this section.

The analysis builds upon Bayes theorem. It gives, for a source with a
measured flux density $S_m$ and uncertainty $\sigma_m$, the posterior
probability distribution of its intrinsic flux density $S_i$: 
\begin{equation}
  \label{eq:6}
  P(S_i \, | \, S_m, \sigma_m ) = \frac{P(S_i) \, P(S_m, \sigma_m \, |
 \,    S_i)}{P(S_m, \sigma_m)},
\end{equation}
where $P(S_i)$ is the probability distribution of intrinsic flux
densities, $P(S_m, \sigma_m \, | \,S_i)$ is the likelihood of the
observed data, and $P(S_m, \sigma_m)$ is a normalization.

The likelihood of observing the data, $P(S_m, \sigma_m \, | \,S_i)$,
is modeled by a Gaussian distribution. This is valid because of the
Gaussian distribution of pixel values in the jackknife maps (see
Fig.~\ref{fig:hist}). The likelihood is
\begin{equation}
  \label{eq:7}
  P(S_m, \sigma_m \, | \,S_i) = \frac{1}{\sqrt{2 \pi \sigma_m^2}} \exp
  \left( {- \frac{(S_m-S_i)^2}{2 \sigma_m^2}} \right)
\end{equation}
We calculated the prior sky distribution $P(S_i)$ from Monte Carlo
simulations. We used a \cite{Schechter:1976aa} form of the number counts
\begin{equation}
  \label{eq:8}
  \frac{d N}{d S} = N' \left( \frac{S}{S'} \right)^{\alpha + 1} \exp{
    \left(- S/S' \right)},
\end{equation}
as fitted to the source counts in the SCUBA SHADES survey of
\cite{CoppinChapin:2006aa}. We used $N' = 1703 \, \mathrm{deg}^{-2} \,
\mathrm{mJy}^{-1}$, $S' = 3.1 \, \mathrm{mJy}$ and $\alpha = -2.0$, which
have been scaled from the \citeauthor{CoppinChapin:2006aa} values
using a spectral index of 2.7. We then drew randomly source fluxes
from that distribution and placed point sources (convolved with the
beam) on noiseless sky maps. The position of each source was drawn
from a uniform distribution, so no clustering was introduced in the
simulation. We simulated a region of 10\arcmin $\times$ 10\arcmin. In
order to reduce edge effects, each simulated map was 3\arcmin~larger,
and a smaller map was extracted.

We generated $\sim 10^6$ sky maps and measured the pixel distribution
in each of them. The mean distribution of pixel values in these maps
is a measure of $P(S_i)$.

From those simulated sky maps, we could calculate the posterior flux
distribution of a source extracted from the map with flux density
$S_m$ and uncertainty $\sigma_m$: it is obtained by multiplying the
prior $P(S_i)$ (from the simulations) with the Gaussian distribution
(eq.~\ref{eq:7}), and dividing by $P(S_m, \sigma_m)$.

Figure~\ref{fig:pfd} shows examples of posterior flux distributions
for three combinations of $S_m$ and $\sigma_m$. The amount of flux
boosting for a certain source depends both on the signal-to-noise
ratio and the value of the flux density; for example, for a
signal-to-noise ratio of 4, a source with a measured flux density of
10~mJy will have been boosted more than a source with a flux density
of 5~mJy.

\end{document}